\documentclass[twocolumn,showpacs,preprintnumbers,amsmath,amssymb]{revtex4}
\usepackage{graphicx}% Include figure files

\usepackage{dcolumn}% Align table columns on decimal point
\usepackage{bm}% bold math
\usepackage{verbatim} % use comment
\usepackage{units}%use nicefrac
\usepackage{tipa}
\usepackage{multirow}
\usepackage{pifont}

\newcommand{\cmark}{\ding{51}}% Checkmark in table of C_n's
\newcommand{\xmark}{\ding{55}}% X in table of C_n's

\newcommand{\be}{\begin{equation}}
\newcommand{\ee}{\end{equation}}
\newcommand{\bea}{\begin{eqnarray}}
\newcommand{\eea}{\end{eqnarray}}

\newcommand{\one}{|1\rangle}
\newcommand{\two}{|2\rangle}

\newcommand{\ut}{U_{\mathrm{trap}}}

\newcommand{\nupa}{\nu_{\mathrm{PA}}}
\newcommand{\ipa}{I_{\mathrm{PA}}}

\newcommand{\Tf}{T_{\mathrm{F}}}

\newcommand{\kb}{k_{\mathrm{B}}}

\begin{document}
\date{\today}
%\flushbottom \draft
\title{High resolution photoassociation spectroscopy of the $^{6}$Li$_2$ $1^{3}\Sigma_{g}^{+}$ state}
\author{Mariusz Semczuk $^1$, Xuan Li$^2$, Will Gunton$^1$, Magnus Haw$^1$, Nikesh S. Dattani$^3$, Julien Witz$^1$, Arthur Mills$^1$, David J. Jones$^1$, and Kirk W.~Madison$^1$}

\affiliation{$^{1}$Department of Physics and Astronomy, University of British Columbia, Vancouver, Canada\\
$^{2}$ Chemical Science Division, Lawrence Berkeley National Laboratory, Berkeley, USA\\
$^{3}$ Department of Chemistry, University of Oxford, Oxford, UK
}

\begin{abstract}
We present experimental observations of seven vibrational levels $v'=20-26$ of the $1^{3}\Sigma_{g}^{+}$ excited state of Li$_2$ molecules by the photoassociation (PA) of a degenerate Fermi gas of  $^6$Li atoms.  For each vibrational level, we resolve the rotational structure using a Feshbach resonance to enhance the PA rates from $p$-wave collisions.   We also, for the first time, determine the spin-spin and spin-rotation interaction constants for this state.  The absolute uncertainty of our measurements is $\pm 0.00002$~cm$^{-1}$ ($\pm 600$ kHz).  We use these new data to further refine an analytic potential for this state.
\end{abstract}


\maketitle

\section{Introduction}

Photoassociation (PA) of ultra-cold atoms is a powerful spectroscopic technique that has been used extensively since the advent of laser cooling to make precise measurements of high-lying vibrational levels that are often difficult to access with traditional bound-bound molecular spectroscopy.  In addition to improving our knowledge of weakly bound molecular states, PA spectroscopy has also allowed precise determinations of atom-atom scattering lengths and excited atomic state lifetimes.  PA resonances have also been used to control atomic interactions via optical Feshbach resonances and for the production of ultra-cold molecules as discussed in several excellent review articles \cite{RevModPhys.71.1, J.Mol.Spectrosc.195, AdvAtMolOptPhys.42, AdvAtMolOptPhys.47, RevModPhys.78.483}.

\begin{figure}[ht]
  \begin{center}
    \includegraphics[width=0.49\textwidth]{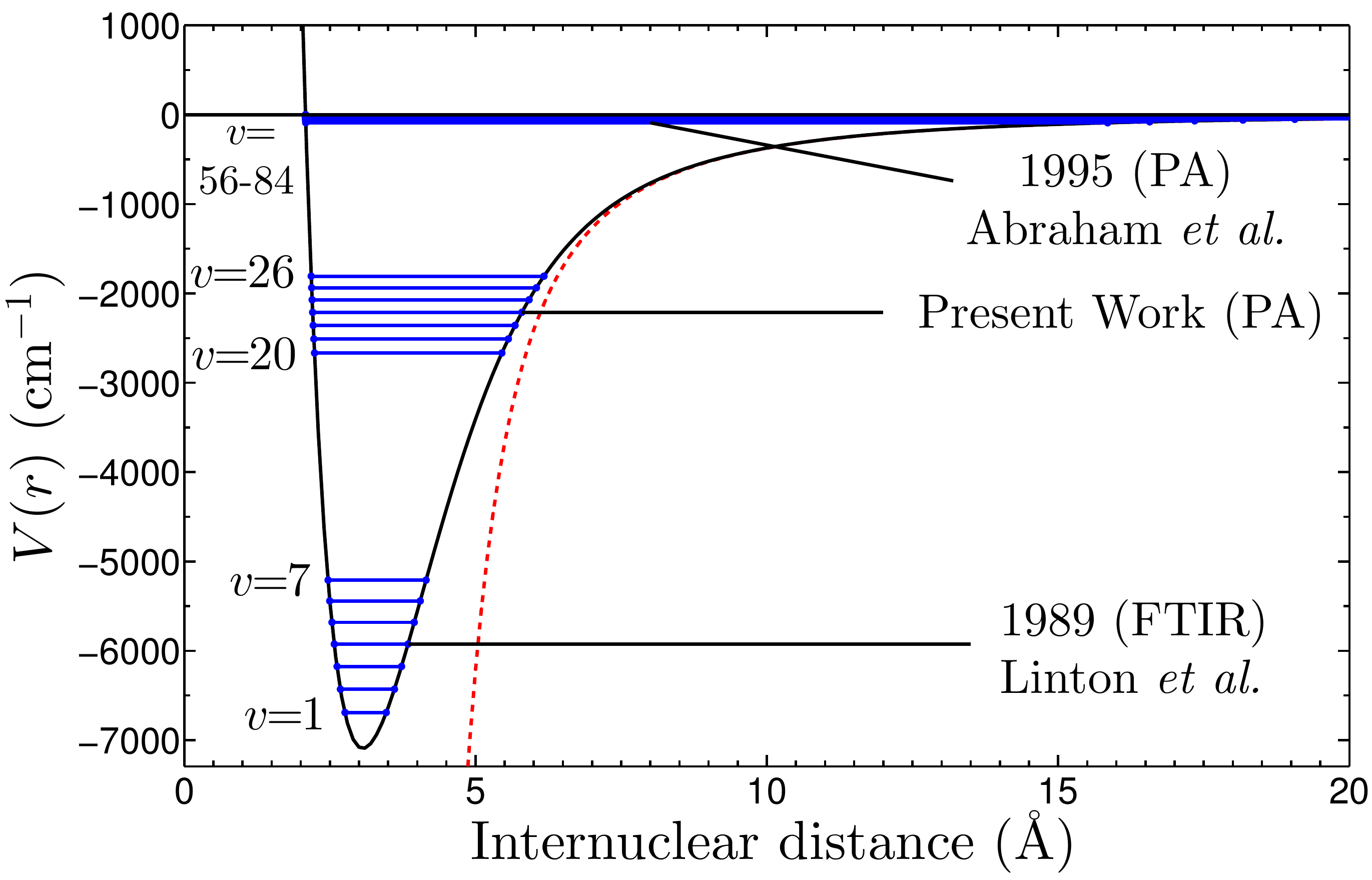}
     \caption{(Color online) The $1^{3}\Sigma_{g}^{+}$ potential studied in this work (solid line).  The solid filled areas indicate
     regions where experimental data is available for $^{6,6}$Li$_2$.  The present work includes high resolution data from seven new 
     vibrational states ($v'=20$ to $26$) including the $N'=0,1,2$ rotational states in each case.  The theoretical long-range potential according to \cite{Aubert-Frecon1998}
     is shown by the dotted line.
     }
  \label{fig:globalPotentialAndDataRegions}
  \end{center}
\end{figure}

In this work, we measure the binding energies of seven vibrational levels $v'=20-26$ of the $1^{3}\Sigma_{g}^{+}$ excited state of $^{6}$Li$_2$ molecules by photoassociating a quantum degenerate Fermi gas of lithium atoms held in a shallow optical dipole trap.  The absolute uncertainty of our measurements is $\pm 0.00002$~cm$^{-1}$ ($\pm 600$ kHz). As in previous high-resolution photoassociative spectra of lithium, our frequency resolution allows us to resolve the rotational structure of these levels \cite{PhysRevA.53.3092}.  In addition, we observe and quantify for the first time ever the spin-spin and spin-rotation coupling constants for this state.  As shown in Fig.~\ref{fig:globalPotentialAndDataRegions}, these measurements are in a completely unexplored spectral range for this molecule and bridge a gap between measurements of the deeply lying $v'=1-7$ levels by Fourier transform spectroscopy (of both $^{7,7}$Li$_2$ and $^{6,6}$Li$_2$ molecules) \cite{Martin19881369,linton:6036} and measurements of the binding energies of levels $v' = 62-90$ of $^{7,7}$Li$_2$ and $v' = 56-84$ of $^{6,6}$Li$_2$ by photoassociation of atoms in a magneto-optic trap \cite{abraham:7773}.

The motivations for the present work of mapping the excited state potential in this wavelength range include the following: (a) as mentioned above, this is a previously unexplored region of the potential and these measurements thus provide an important addition to the existing data allowing us to make a much more complete and accurate, global description of this state, (b) knowing the locations of these intermediate states is important for the eventual formation of ultracold ground triplet state molecules since these states are expected to strike the best compromise between a good Franck--Condon overlap with the initial state (either a Feshbach resonance molecule or an unbound collision state) and the final state in the $a^3\Sigma_u^{+}$ potential, (c) this wavelength range is particularly convenient for future experiments since it is easily accessible by both solid-state (Ti:sapphire) and diode lasers.  This latter point is relevant to future experiments on the probing, alignment and spinning of ultracold, weakly bound Li$_2$ molecules (i.e.~either Feshbach halo dimers or molecules in the ground triplet state) with high-intensity, ultra-short pulses from Ti:sapphire lasers \cite{deiglmayr:064309,PhysRevA.79.050501,PhysRevLett.103.053003}.  In such experiments, knowing the excited state energy levels is important as off resonant excitation of the ground state molecule is intended and excited state transfer should be avoided.

This paper is organized as follows.  Section \ref{sec:exp} presents a description of the experimental apparatus and our measurement methodology.  In Sec.~\ref{sec:mes} we show several examples of measured spectra and in Sec.~\ref{sec:int} we discuss the interpretation and assignment of the PA spectral features.  In Sec.~\ref{sec:thy}, we introduce the model for the potential energy curve for Li$_2$ ($1^3\Sigma_g^+$), and we present refined analytic potentials for this state and the $a^3\Sigma_g^+$ state of $^{6,6}$Li$_2$  and $^{7,7}$Li$_2$, that were calculated using experimental results from this paper.  We conclude with a summary and outlook for future work in Sec.~\ref{sec:conc}.

\section{Experimental Methods}
\label{sec:exp}

For these measurements, we load a magneto-optic trap (MOT) with $3 \times 10^{7}$ $^{6}$Li atoms in 20 seconds directly from an effusive oven as described in \cite{Ladouceur:09, PhysRevA.82.020701}. We then compress and cool the MOT by increasing the axial magnetic field gradient from $40$ to $60$~G cm$^{-1}$, lowering the intensity and shifting the frequency of both the ``pump" light (near the $2s_{1/2}, \, F=3/2 \rightarrow 2p_{3/2}, \, F=5/2$ transition frequency) and the ``repump" light (near the $2s_{1/2}, \, F=1/2 \rightarrow 2p_{3/2}, \, F=3/2$ transition frequency) to 10~MHz below resonance.  During this compression and cooling phase, a crossed dipole trap (CDT) of 200 W total power is turned on and, in less than 10~ms, 5\% of the $^{6}$Li atoms are transferred into the CDT.  We observe extremely rapid trap losses due to light assisted collisions and hyperfine relaxation, and we therefore optically pump to the lower hyperfine state ($F=1/2$) during the transfer by extinguishing the ``repump" light.  This procedure produces an almost equal population of the two sub-levels of the lower hyperfine state : $\one \equiv |F=1/2, m_F=1/2\rangle$ and $\two \equiv |F=1/2, m_F=-1/2\rangle$.  The light for the CDT is derived from a 100~W fiber laser (SPI Lasers, SP-100C-0013) operating at 1090 nm with a spectral width exceeding 1 nm.  The CDT is comprised of two nearly co-propagating beams crossing at an angle of 14$^{\circ}$.  Each beam has a maximum power of 100~W (for a total power of 200 W) and is focused to a waist ($1/e^2$ intensity radius) of $42 \, \mu$m and $49 \, \mu$m.  After the MOT light is extinguished, the CDT beam power is ramped down linearly in time to 100 W total (50 W per beam) in 100~ms while applying a homogenous magnetic field of 800~G.  Rapid thermalization occurs because of the large collision rate induced by a very wide $s$-wave Feshbach resonance between the $\one$ and $\two$ states at 834~G \cite{PhysRevA.71.045601}.  At the end of this forced evaporation stage, there are approximately $10^6$ atoms remaining at a temperature of $200 \, \mu$K (verified by a time-of-flight expansion measurement).  At this point, atoms are transferred into a lower-power CDT which is superimposed on the high-power CDT.  The light for this second CDT is generated by a narrow-linewidth ($< 10$ kHz), 20 W fiber laser operating at 1064~nm (IPG Photonics YLR-20-1064-LP-SF).  This transfer is done to avoid ensemble heating observed in the SPI laser CDT and allows further forced evaporative cooling to much lower temperatures.  The IPG CDT is comprised of two beams crossing at an angle of 60$^{\circ}$ and with a total power of 15 W.  The beams are focused to a waist ($1/e^2$ intensity radius) of $25 \, \mu$m and $36 \, \mu$m.  The IPG CDT beam powers are controlled by two independent acousto-optic modulators and are configured to have a frequency difference of 190~MHz so that the rapidly moving interference pattern generated at the intersection of the two beams has no effect on the atomic motion and the atoms experience the averaged potential of the two beams.  After the transfer of atoms into the IPG CDT, the magnetic field is lowered to 300~G, where the $s$-wave collision cross section between the $\one$ and $\two$ states is large enough to continue efficient evaporation.  This magnetic field is also chosen because at this value there are no bound molecular states near to the atom-atom threshold into which the atoms can decay via three-body recombination.  The goal of this is to produce very cold atomic distributions without also forming Feshbach molecules which would occur for evaporation near the 834~G Feshbach resonance.  The trap depth is lowered using a combination of linear and exponential ramps from $\ut=500 \, \mu$K to $\ut=8 \, \mu$K in typically 5~s.  At the end of this evaporation step, the ensemble is composed of $3 \times 10^4$ atoms with equal populations of the $\one$ and $\two$ states, and the temperature (verified by time-of-flight expansion) is $800$~nK.

After this preparation step, the magnetic field is lowered to a very small value and PA light from a single-frequency, tunable Ti:sapphire laser beam illuminates the atomic cloud for a certain exposure time.  The PA light is a single beam that propagates co-linearly with one of the arms of the lower-power CDT and is focused to a waist ($1/e^2$ intensity radius) of 50~$\mu$m.  The light is linearly polarized and aligned along the direction of the bias magnetic field used for the measurements of $p$-wave Feshbach enhanced PA.  When the bias field is off, there persists a residual magnetic field below 400~mG.  For these experiments, the power of the PA light is up to 100 mW corresponding to an intensity of 1270~W~cm$^{-2}$.  When the photon energy $h \nupa$ equals the energy difference between the unbound state of a colliding atomic pair and a bound molecular excited state, molecules form at a rate proportional to the atom-atom collision rate and atoms are subsequently lost from the trap.  This loss occurs because the excited state molecule either radiatively decays into the unbound continuum of two free atoms with sufficient energy to be lost from the shallow CDT or it decays into a bound state molecule which is not detected in our atom number measurement.  The probability of this latter event can be quite high when exciting particular vibrational levels in the $1^{3}\Sigma_{g}^{+}$ excited state \cite{cote1999}.  After this exposure time, the number of atoms remaining is determined by an absorption image of the cloud immediately after the extinction of the CDT.

For some of the measurements presented here, $\nupa$ was determined by a commercial wavemeter (Bristol 621A-NIR) with an absolute accuracy of 60~MHz and a shot-to-shot repeatability (i.e.~precision) of 10 MHz in the frequency range of this work.  For the high resolution measurements, the Ti:sapphire laser, operating in the range from 770 to 820 nm, is stabilized to a fiber based, self-referenced frequency comb operating with a center wavelength of 1550 nm as described previously \cite{Mills:09}.  Briefly, the frequency comb is an Erbium-doped fiber laser frequency comb with two amplified output branches.  One branch is used for self-referencing the carrier-envelope offset frequency via an $f$--$2f$ interferometer.  The second branch is also spectrally broadened in a highly nonlinear fiber, but not to a full octave of optical frequencies.  The output of this branch is frequency-doubled using an array of periodically-poled lithium niobate waveguides with different poling periods. The frequency-doubled comb is then mixed with the Ti:sapphire laser on a fast photodiode to generate a heterodyne beatnote, which is used to stabilize the Ti:sapphire laser to the frequency comb.  For this work we verified the comb-referenced Ti:sapphire's absolute frequency uncertainty by measuring the resonant frequencies of the D2 line at 780~nm (the $5s_{1/2}, \rightarrow 5p_{3/2}$ transition) of $^{85}$Rb atoms in a vapor cell and comparing them with their known values \cite{steck85}.  We verified that the absolute uncertainty is $\pm600$~kHz, consistent with that determined previously \cite{Mills:09}.

\section{Observations}
\label{sec:mes}

\begin{figure}[ht]
  \begin{center}
    \includegraphics[width=0.49\textwidth,angle=0]{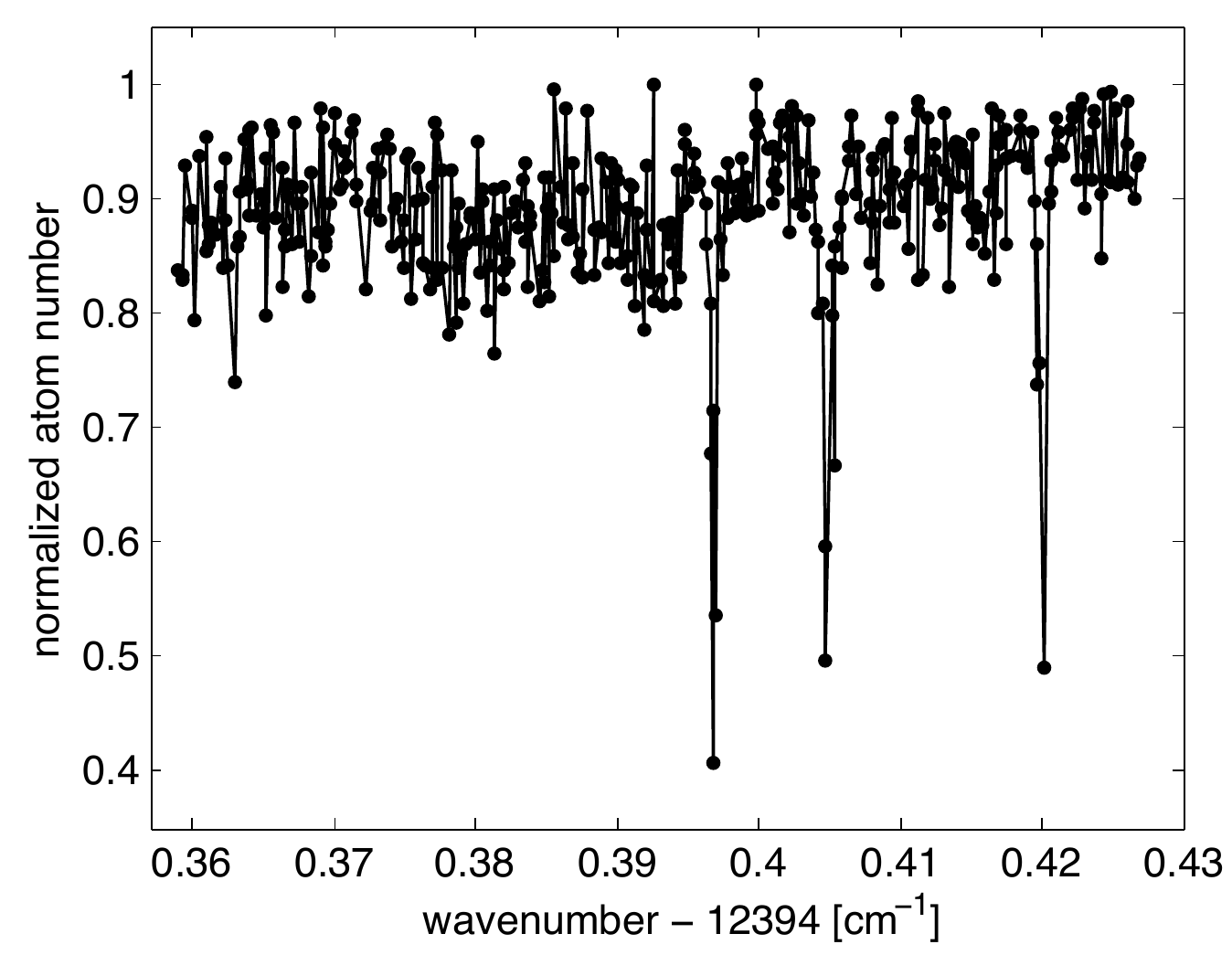}
     \caption{Normalized $^{6}$Li atom number as a function of photo-association laser energy $h \nupa$ after a 2 second hold time with
zero bias magnetic field and a PA laser intensity of $\ipa = 635$~W~cm$^{-2}$.  These three resonances correspond to a transition from an initial unbound molecular state with $N=0, G=0$ to the $v'=21$ vibrational level of the $1^{3}\Sigma_{g}^{+}$ excited state with $N'=1, G'=0$.  The ensemble temperature was 15~$\mu$K.
     }
  \label{fig:three-features}
  \end{center}
\end{figure}

\begin{figure}[ht]
 \begin{center}
  \includegraphics[width=0.49\textwidth,angle=0]{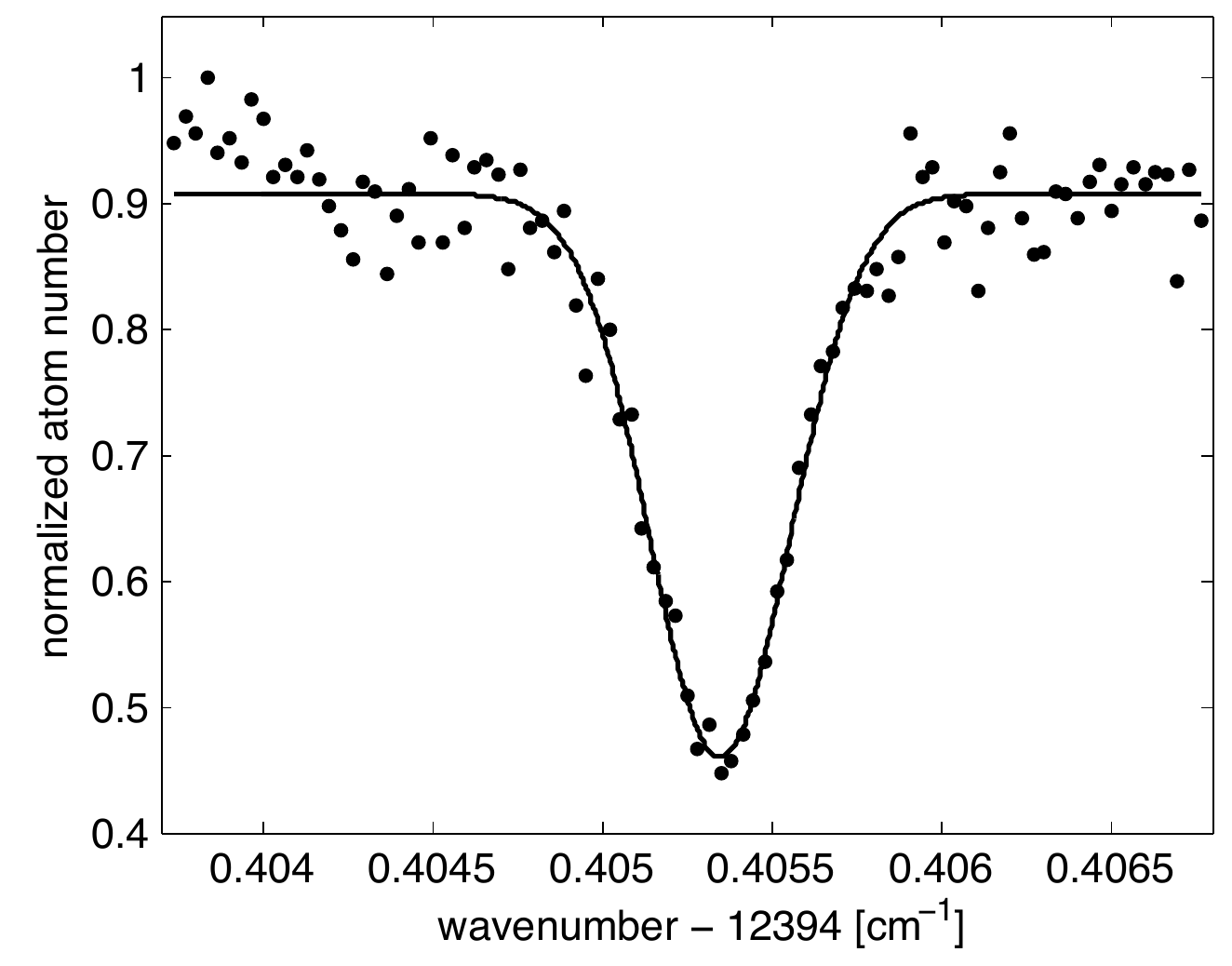}
     \caption{High resolution scan of the normalized $^{6}$Li atom number as a function of photo-association laser energy $h \nupa$ after a 750~ms hold time with
zero bias magnetic field and a PA laser intensity of $\ipa = 635$~W~cm$^{-2}$.  This is the second of the three resonances shown in Fig.~\ref{fig:three-features} corresponding to a transition from an initial unbound molecular state with $N=0, G=0$ to the $v'=21$ vibrational level of the $1^{3}\Sigma_{g}^{+}$ excited state with $N'=1, G'=0$.  The ensemble temperature was 800~nK.  The FWHM of this loss peak is $0.00048$~cm$^{-1}$ (14.4 MHz).
     }
  \label{fig:high-res-feature}
  \end{center}
 \end{figure}

In our initial search of PA resonances, we held the magnetic field near to the $s$-wave Feshbach resonance between the  $\one$ and $\two$ states at 834~G to enhance the collision rate.  This produced a very wide (1~GHz FWHM) PA loss feature which facilitated initial detection \cite{MagnusThesis}.  After the approximate locations of the PA resonances were found in this way, we performed a high resolution scan with an ensemble temperature of 15~$\mu$K and with no bias magnetic field \footnote{We verified that the magnetic field was below 400~mG by measuring the transition frequencies between the magnetic sublevels of the F=2 and F=3 ground hyperfine states of $^{85}$Rb prepared in the ODT.}.  We observed that the PA spectrum of each vibrational level had associated with it three narrow (below 10 MHz FWHM) features distributed across a range of 0.7~GHz as shown in Fig.~\ref{fig:three-features}.  Figure~\ref{fig:high-res-feature} shows a higher resolution scan of the second feature shown in Fig.~\ref{fig:three-features}.  In order to reduce as much as possible the thermal broadening and the inhomogeneous AC Stark shift produced by the optical dipole trapping potential, these data were obtained in a very shallow trap ($\ut / \kb \sim 8 \, \mu$K) and an ensemble temperature of $800$~nK, a temperature well below the Fermi temperature for this two component Fermi-gas ($T/\Tf = 0.4$).  We then verified that these PA resonances arise from collisions between atoms in states $\one$ and $\two$ by using a state-selective resonant pulse of light to remove all atoms in either of the two states.  The spin purification was done at the end of the preparation sequence, and we observed the absence of these atom loss features with either one of the states removed \footnote{The spin purification was performed at a high magnetic field (typically 700 G), where the optical transitions from the  $\one$ and $\two$ states to the excited  $2p_{3/2}$ manifold are well separated.  In addition, the field is sufficiently large to disrupt the hyperfine coupling and these transitions become approximately ``closed" such that the excited state atom returns to the original ground state with a large probability allowing each atom to scatter many photons during the pulse and subsequently leave the trap.}. To rule out the absence of these loss features due to a simple reduction of the density, we observed a reappearance of the PA features when using an incoherent mixture of the $\one$ and $\two$ states with the same total number of particles and temperature as the ensembles after spin purification.  Given that $p$-wave collisions are dramatically suppressed at these temperatures and that these PA loss features were visibly enhanced by the $s$-wave FR, we inferred that they arise from $s$-wave collisions between atoms in states $\one$ and $\two$.  Thus, they correspond to a transition from an initial unbound molecular state with $N=0, G=0$ to an excited state with $N'=1, G'=0$ (assuming $G$ is a good quantum number).  As we describe later, we find that spin-spin and spin-rotation coupling split the excited state into three sub-levels producing the three PA features.  In this case $G$ is no longer a good quantum number.  The locations of these three features for each of the seven vibrational levels is provided in Table~\ref{tab:N1-resonances}.

\begin{figure}[ht]
  \begin{center}
    \includegraphics[width=0.49\textwidth,angle=0]{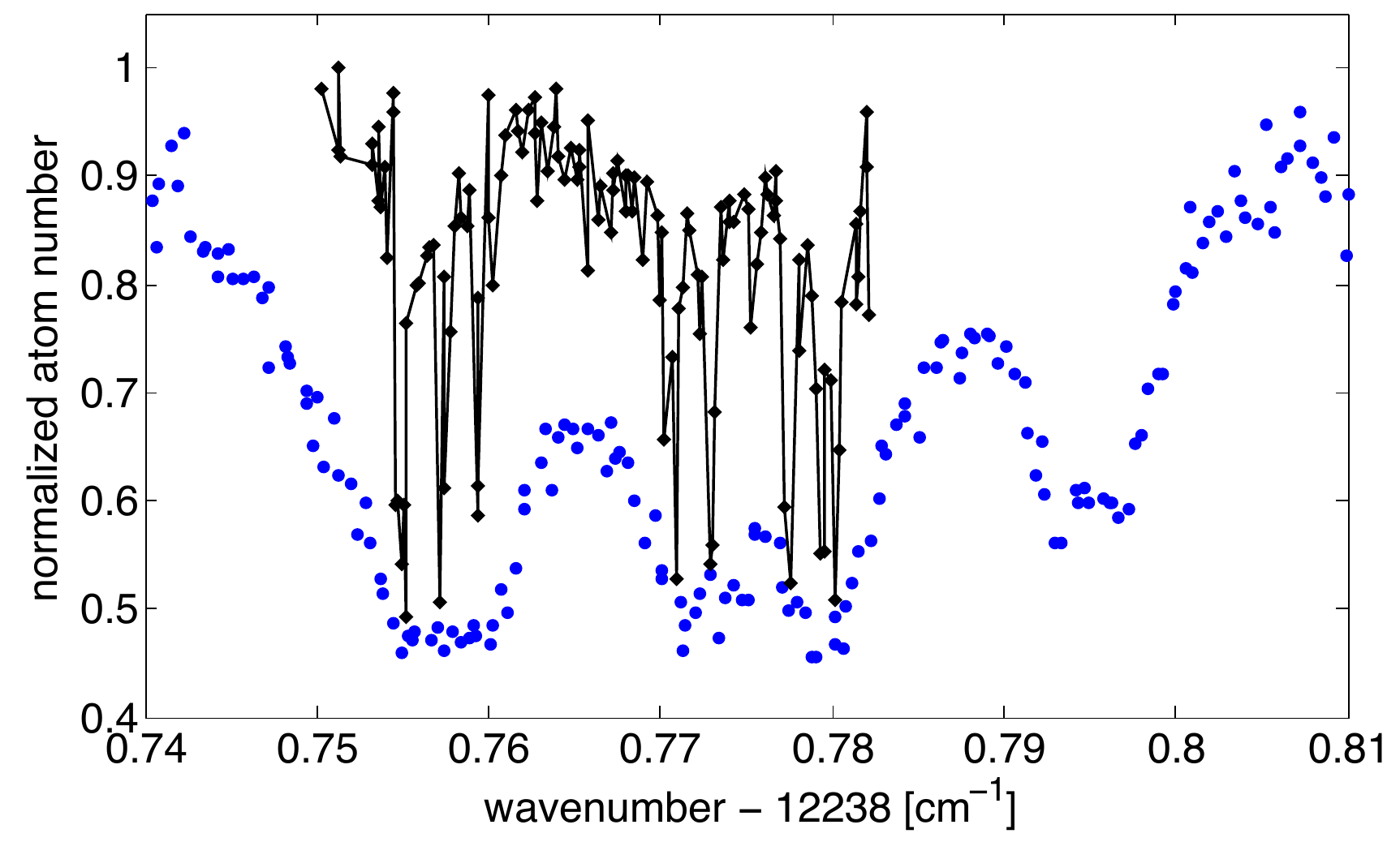}
     \caption{(Color online) Normalized $^{6}$Li atom number as a function of photo-association laser energy $h \nupa$ after a 2 second hold time.  The circles are for an ensemble temperature of 250~$\mu$K at 185~G, and four distinct features are observed.  The diamonds denote the atom loss for an ensemble temperature of 15~$\mu$K and at a magnetic field of 184.7~G.  At this lower temperature, these loss features are seen to result from multiple PA resonances that are unresolvable at 250~$\mu$K. These PA features arise from $p$-wave ground-state collisions and are enhanced by proximity to a $p$-wave Feshbach resonance between the $\one$ and $\two$ states at 185.1~G.}
  \label{fig:pwave-feature}
  \end{center}
\end{figure}

We also located for each of the vibrational states the PA resonances associated with $p$-wave ground-state collisions.  However, these features were \emph{only} observable in our experiment when measures were taken to enhance the PA scattering rate.  In order to observe these PA resonances, we enhanced the $p$-wave scattering rate by stopping the evaporation at an ensemble temperature of 250~$\mu$K and by holding the magnetic field at 185~G during the PA stage.  This magnetic field is near the $p$-wave Feshbach resonance between the $\one$ and $\two$ states at 185.1~G \cite{PhysRevA.71.045601}.  Due to the Feshbach resonance enhancement of inelastic ground-state collisions, the ensemble particle loss in the absence of the PA light was approximately 50\% during the 2 second hold time.  Additional loss was induced when the light was near a PA resonance.  Figure~\ref{fig:pwave-feature} shows the loss spectrum for a transition from an initial unbound molecular state with $N=1, G=1$ to the $v'=20$ vibrational level of the $1^{3}\Sigma_{g}^{+}$ excited state with $N'=2, G'=1$.  For each of the seven vibrational levels, we observed at least 4 (3) distinct loss features for transitions to the $N'=2, G'=1$ ($N'=0, G'=1$) final state.  By evaporating the ensemble to 15~$\mu$K and holding the magnetic field at 184.7~G, we observed that each of these loss features results from multiple PA resonances that are unresolvable at 250~$\mu$K.  The locations of the loss features observed at 250~$\mu$K for each of the seven vibrational levels is provided in Tables~\ref{tab:N0-resonances} and \ref{tab:N2-resonances}.  These measurements were performed in the absence of the comb stabilization.  Instead, the Ti:sapphire laser was referenced to the wavemeter whose uncertainty is 60 MHz.

\begin{table}
\caption{Experimentally measured PA resonances for $s$-wave collisions in an incoherent mixture of the $\one$ and $\two$ states of $^6$Li.  These three PA resonances correspond to a transition from an initial unbound molecular state with $N=0, G=0$ to the $v^{\mathrm{th}}$ vibrational level of the $1^{3}\Sigma_{g}^{+}$ excited state with $N'=1$.  As we explain in Sec.~\ref{sec:int}, spin-spin and spin-rotation coupling split the excited state into three sub-levels producing the three PA features corresponding to quantum numbers $(N'=1, J'=1)$, $(N'=1, J'=2)$, and $(N'=1,  J'=0)$ respectively.  The absolute uncertainty in each these measurements is  $\pm 0.00002$ cm$^{-1}$ ($\pm 600$kHz).
}
\begin{ruledtabular}
\begin{tabular}{c c c c}

$v'$ & 1st & 2nd & 3rd \\
 & cm$^{-1}$ &  cm$^{-1}$ &  cm$^{-1}$ \\
  \hline
20 & 12237.17755 & 12237.18587 & 12237.20126\\
  \hline
21 & 12394.39726 & 12394.40535 & 12394.42039\\
  \hline
22 & 12546.06767 & 12546.07552 & 12546.09025\\
  \hline
23 & 12692.17316 & 12692.18080 & 12692.19509\\
  \hline
24 & 12832.70080 & 12832.70820 & 12832.72214\\
  \hline
25 & 12967.64150 & 12967.64862 & 12967.66219\\
  \hline
26 & 13096.99114 & 13096.99804 & 13097.01125\\
\end{tabular}
\end{ruledtabular}
\label{tab:N1-resonances}
\end{table}

\begin{table}
\caption{Experimentally measured PA resonances for $p$-wave collisions in an incoherent mixture of the $\one$ and $\two$ states of $^6$Li held at a magnetic field of $B=185$~G.  Each of these values was extracted by fitting a loss spectrum like that shown in Fig.~\ref{fig:pwave-feature}.  These PA resonances correspond to a transition from an initial unbound molecular state with $N=1, G=1$ to the $v^{\mathrm{th}}$ vibrational level of the $1^{3}\Sigma_{g}^{+}$ excited state with $N'=0, G'=1$.  While the precision in these measurements is 0.001 cm$^{-1}$, the uncertainty, limited by the wavemeter, is $\pm0.002$ cm$^{-1}$ 
}
\begin{ruledtabular}
\begin{tabular}{c c c c }

$v'$ & 1st & 2nd & 3rd \\
 & cm$^{-1}$ &  cm$^{-1}$ &  cm$^{-1}$ \\
  \hline
20 & 12236.388 & 12236.407 & 12236.424\\
  \hline
21 & 12393.629 & 12393.648 & 12393.664\\
  \hline
22 & 12545.320 & 12545.338 & 12545.355\\
  \hline
23 & 12691.446 &  12691.465 & 12691.480\\
  \hline
24 & 12831.995 & 12832.012 & 12832.029\\
  \hline
25 & 12966.957 & 12966.975 & 12966.991\\
  \hline
26 & 13096.326 & 13096.346 & 13096.362\\
\end{tabular}
\end{ruledtabular}
\label{tab:N0-resonances}
\end{table}

\begin{table}
\caption{Experimentally measured PA resonances for $p$-wave collisions in an incoherent mixture of the $\one$ and $\two$ states of $^6$Li held at a magnetic field of $B=185$~G.  Each of these values was extracted by fitting a loss spectrum like that shown in Fig.~\ref{fig:pwave-feature}.  These PA resonances correspond to a transition from an initial unbound molecular state with $N=1, G=1$ to the $v^{\mathrm{th}}$ vibrational level of the $1^{3}\Sigma_{g}^{+}$ excited state with $N'=2, G'=1$.   While the precision in these measurements is 0.001  cm$^{-1}$, the uncertainty, limited by the wavemeter, is $\pm0.002$ cm$^{-1}$}
\begin{ruledtabular}
\begin{tabular}{c c c c c }

$v'$ & 1st & 2nd & 3rd & 4th \\
 & cm$^{-1}$ &  cm$^{-1}$ &  cm$^{-1}$ &  cm$^{-1}$ \\
  \hline
20 & 12238.757 & 12238.772 & 12238.780 & 12238.795 \\
  \hline
21 & 12395.936 & 12395.951 &12395.958 & 12395.973\\
  \hline
22 & 12547.567 & 12547.579 & 12547.587 & 12547.601\\
  \hline
23 & 12693.628 & 12693.642 & 12693.648 & 12693.665\\
  \hline
24 & 12834.113 & 12834.128 & 12834.134 & 12834.150\\
  \hline
25 & 12969.011 & 12969.026 & 12969.032 & 12969.047\\
  \hline
26 & 13098.315 & 13098.332 & 13098.339 & 13098.355\\

\end{tabular}
\end{ruledtabular}
\label{tab:N2-resonances}
\end{table}

\subsection{Characterization of systematic shifts}

While the absolute uncertainty of our PA measurements made using the frequency comb is $\pm600$~kHz, the data was taken in the presence of a small but non-zero magnetic field and in an optical dipole trap with a known intensity.  These residual fields as well as the PA laser itself can lead to a systematic shift of the resonance positions from their zero-field values. Therefore, in an effort to quantify the role of the PA laser intensity, the CDT laser intensity, and the residual magnetic field on the PA loss features, we varied each one and measured the PA resonance position and width for various excited states.  In each case, we assumed a linear dependence and determined a shift rate of the resonance position with the corresponding field strength.  The uncertainty in this rate is a one-sigma statistical uncertainty on the slope of the linear fit.

When varying the PA laser intensity from $\ipa = 0.19$~kW~cm$^{-2}$ to $\ipa = 1.27$~kW~cm$^{-2}$ we observed that the centroid of the first feature ($J'=1$) associated with the $v'=26$ excited state shifted to higher frequencies at a rate of $471 \pm 433$~kHz per kW cm$^{-2}$.  When the CDT laser intensity was varied from 5.4~kW~cm$^{-2}$ (145 mW total CDT power) to 140~kW~cm$^{-2}$ (3.1 W total CDT power) the PA feature centroid associated with the $v'=24, J'=1$ state shifted down in frequency at a rate of $-(19 \pm 1.2)$~kHz per kW cm$^{-2}$.  The resonance positions reported in Table~\ref{tab:N1-resonances} were determined using a PA laser intensity of $\ipa = 635$~W~cm$^{-2}$, and a CDT intensity of 7.5~kW~cm$^{-2}$.  Assuming the differential AC Stark shift is the same for all excited states, the reported values are therefore shifted lower by $142 \pm 9$~kHz due to the CDT and higher by $300 \pm 274$~kHz due to the PA laser than their extrapolated position at zero differential AC Stark shift.  The overall AC Stark shift of the resonance positions is thus higher by $157$~kHz with an uncertainty of $\pm 274$~kHz.  Both this shift and uncertainty are small compared to the absolute uncertainty of the frequency comb.  For the resonance positions reported in Tables~\ref{tab:N0-resonances} and \ref{tab:N2-resonances}, the trapping power was larger (40~W total) and the differential AC Stark shift due to the CDT is estimated to be $-(15 \pm 1)$~MHz. 

When the magnetic field was varied from 0~G to 10~G the PA features associated with the $v'=24, J'=1$, $J'=2$, and $J'=0$ states were observed to shift and, in the case of $J'=1$ and $J'=2$, to broaden and eventually split into multiple resolvable peaks.  In each case, we measured the PA feature center of mass and found that when the magnetic field was varied from 0 to 1 G, the barycenter of the PA features moved by $-(91.2 \pm 18.3)$~kHz for the $J'=1$ state, $+(46 \pm 28)$~kHz for the $J'=2$ state, and $+(74.5 \pm 30.1)$~kHz for the $J'=0$ state.  Since the resonance positions reported in Table~\ref{tab:N1-resonances} were determined in the presence of a residual magnetic field below 400~mG, the uncertainty in their positions due to the magnetic field was below 50~kHz for all $J$ states and thus small compared to the absolute uncertainty of the frequency comb.

\section{Interpretation}
\label{sec:int}

In order to interpret our measurements, we begin with a brief review of the symmetry properties and corresponding selection rules relevant for the photoassociation process.  Molecules in the $1^{3}\Sigma_{g}^{+}$ excited state are characterized by the Hund's case ``b" coupling scheme in which the total electronic (nuclear) spin $\vec{S} = \vec{s}_1 + \vec{s}_2$ ($\vec{I} = \vec{i}_1 + \vec{i}_2$) is completely uncoupled from the internuclear axis.  Here $\vec{s}_j$ ($\vec{i}_j$) is the electronic (nuclear) spin of atom ``$j$".  This occurs when $\Lambda = 0$, the projection of the orbital angular momentum of the electrons along the internuclear axis is zero, and there is therefore no axial magnetic field to couple the total spin to the axis.  For ``$\Sigma$" states, the orbital angular momentum of the electrons is zero and therefore $\Lambda$ is always identically zero;  however, even in some cases where $\Lambda \ne 0$, especially for light molecules, the coupling is sufficiently weak that Hund's case ``b" is still the appropriate scheme \cite{Herzberg}.  The total angular momentum, apart from the spin, is $\vec{K}\equiv \vec{N} + \vec{\Lambda}$, the vector sum of $\vec{\Lambda}$ and the rotational angular momentum of the nuclei $\vec{N}$.  Therefore for ``$\Sigma$" states $\vec{K} = \vec{N}$, and thus $\vec{K}$ is perpendicular to the internuclear axis.  The total spin of the molecule is $\vec{G} = \vec{S} + \vec{I}$ and is a good quantum number so long as the hyperfine interaction and spin-rotational couplings are small. The total spin combines with the total angular momentum apart from spin $\vec{K}$ to result in the total angular momentum including spin as $\vec{J} = \vec{K} + \vec{G}$.

For electric dipole radiation, the selection rule is that $\Delta J = 0, \pm 1$ with the restriction that $J=0 \nleftrightarrow J=0$.  In addition, under the emission or absorption of a photon the parity of the electronic orbital must change  ($+ \leftrightarrow -$) and for a homonuclear molecule, the symmetry of the coordinate function under interchange of the two nuclei must change from symmetric to anti-symmetric or vice versa ($g \leftrightarrow u$).  In the present scenario of Hund's case ``b" coupling, the spin is so weakly coupled to the other angular momenta that both quantum numbers $S$ and $K$ are well defined and we have in addition the selection rules $\Delta S=0$ (or equivalently $\Delta G =0$) and therefore $\Delta K = 0, \pm 1$
with the restriction that $\Delta K=0$ is forbidden for $\Sigma \rightarrow \Sigma$ transitions.  Since we are here only concerned with transitions to the $1^{3}\Sigma_{g}^{+}$ excited state, we have that $\Delta N = \pm 1$ and $\Delta G = 0$.

\begin{table}[h]
\caption{Allowed rotational levels and corresponding nuclear spin configurations for $^{6}\mathrm{Li}_2$ molecules in the limit that spin-spin and spin-rotation couplings are small enough that $G$ is a good quantum number.}
\begin{tabular}{c l l l l}
State &  Electronic & Nuclear & Allowed & Total \\
 &  spin & spin & rotational states &  Spin \\
 \hline
 \hline
 \multicolumn{5}{c}{ground states}\\
 \hline

\hline
 - & - & - & $N=0,2,4\ldots$ & $G=0$\\
 - & - & - & $N=1,3,5\ldots$ & $G=1$\\
 \hline

 \hline
 \hline
 \multicolumn{5}{c}{excited states}\\
 \hline

 $\mathrm{1}^{3}\Sigma_{g}^{+}$ : & $S=1$ & $I=0$ & $N=0,2,4\ldots$ & $G=1$\\
 & & $I=1$ & $N=1,3,5\ldots$ & $G=0,1,2$\\
 & & $I=2$ & $N=0,2,4\ldots$ & $G=1,2,3$\\
 \hline

\end{tabular}
\label{tab:allowed}
\end{table}

We now discuss the allowed quantum numbers for the initial and final states.  In this work, we only consider collisions between two $^6$Li atoms, which are composite fermions (consisting of 9 fermions: 3 protons, 3 neutrons, and 3 electrons), and we note that the 2-body eigenstates, composed of a spin part and an orbital part, must be antisymmetric upon exchange of the two atoms.  The consequence is that only certain spin states are possible given a particular orbital state.  An important example of this constraint imposed by exchange symmetry is that the two-body position wave function (sometimes called the ``coordinate function" or orbital state) must be antisymmetric for a collision between two fermions in the same spin state (for which the spin wave function is manifestly symmetric).  Thus a spin polarized Fermi gas can only have odd partial wave collisions ($p$-, $f$-, $h$-wave, etc...) corresponding to odd values of the rotational angular momentum of the complex ($N = 1, 3, 5 \ldots$), which are antisymmetric with respect to atom exchange. For a gas composed of two distinct spin states, even partial wave collisions can occur ($s$-, $d$-, $g$-wave, etc...) so long as the spin wave function is antisymmetric upon atom exchange.  As we described in Sec.~\ref{sec:mes}, the ability to turn off $s$-wave collisions by spin polarizing the gas is a useful feature of our system that we use to validate our assignment of the PA lines.

The total spin angular momentum of the initial unbound molecular state is given by the vector sum of the $f$ quantum numbers for the isolated atoms: $\vec{G} = \vec{f}_1 + \vec{f}_2$.  Here $\vec{f}_1 = \vec{s}_1 + \vec{i}_1$.  In our experiment, the atoms are optically pumped to the lowest hyperfine state before being exposed to the photoassociation light.  Therefore we have that $f_1=f_2=\frac{1}{2}$ and there are two allowed values of the total spin: $G=0,1$.  Certain values of $G$ (specifically $G = f_1+f_2, f_1+f_2-2,\ldots$) are associated with spin states symmetric with respect to interchange of the atoms while the orbital states with even values of $N$ are symmetric under the interchange of the atoms.  Therefore all even partial wave collisions ($N=0,2,4,\ldots$) have a total spin of zero ($G=0$) and all odd partial wave collisions ($N=1,3,5,\ldots$) have a total spin of one ($G=1$).

The final state is a molecule in the $1^{3}\Sigma_{g}^{+}$ potential.  For this triplet state, the total electronic spin is well defined ($S=1$) and the ``gerade" symmetry signified by a sub-script ``$g$" denotes that all states with an even rotational quantum number ($N = 0, 2, 4, \ldots$) are symmetric under the interchange of the two nuclei.  Because the electronic spin is well defined and fixed for this excited state, we now consider interchanging just the \emph{nuclei} while leaving the electrons untouched.  There are three possible values of the total nuclear spin ($I=0,1,2$) since the nuclear spin of each atom is $i=1$.  Similar to the symmetry of $G$, states with $I = i_1 + i_2, i_1 + i_2 - 2 \ldots$ (corresponding here to $I=0$ and $I=2$) are symmetric with respect to interchange of the nuclei whereas the $I=1$ state is antisymmetric.  Since the nuclei are \emph{bosons} the total wave function must be symmetric under the interchange of the nuclei.  Putting this together, we have that the even (odd) values of $I$ occur with even (odd) values of $N$.  The total spin angular momentum quantum number $G$ can take on all values between and including $|I+S|$ and $|I-S|$.

\begin{figure}[ht]
  \begin{center}
    \includegraphics[width=0.49\textwidth]{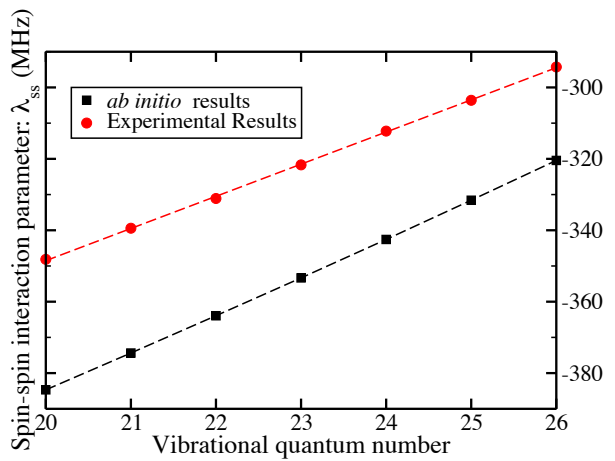}
     \caption{(Color online) The experimentally determined (circles) and {\it ab initio} computed (squares) spin-spin interaction constants, $\lambda_v$, as a function of the vibrational quantum number for the $1^{3}\Sigma_{g}^{+}$ electronic state.  These constants were determined from the frequency splittings of the three features observed for the $N=0 \rightarrow N'=1$ transition. The uncertainty in these values is $\pm 400$~kHz.  The dashed lines are guides to the eye.
     }
  \label{fig:spinspin}
  \end{center}
\end{figure}

The possible quantum numbers for the ground and excited states are tabulated in Table~\ref{tab:allowed}.  For a ground state $s$-wave collision ($N=0$) we find that there is only one allowed value for the total spin: $G=0$.  From an initial state with $N=0$ and $G=0$, we see that there is only one possible transition to the excited triplet state: $(N=0, G=0) \rightarrow (N'=1, G'=0)$.  For a ground state $p$-wave collision, the initial state is $(N=1, G=1)$ and there are two possible transitions to the excited triplet state: $(N=1, G=1) \rightarrow (N'=0, G'=1)$ and $(N=1, G=1) \rightarrow (N'=2, G'=1)$.  In both cases, there are two possible values of the total nuclear spin: $I=0$ or $2$.

\begin{table}
\caption{The values for the spin-spin interaction constant, $\lambda_v$, and the spin-rotation interaction constant, $\gamma_v$, determined from Eq.~\ref{SplittingEq2} and the peak spacings reported in Table \ref{tab:N1-resonances}. The uncertainty in these values is $\pm 400$~kHz.  The $\lambda_v$ values are plotted in Fig.~\ref{fig:spinspin} along with their expected values determined from {\it ab initio} calculations.
}
\begin{ruledtabular}
\begin{tabular}{c c c }

$v'$ & $\lambda_v$ (MHz)  & $\gamma_v $ (MHz) \\
\hline
20 & -348.2 & -14.5 \\
\hline
21 &  -339.4 & -14.5 \\
\hline
22 &  -331.1 & -14.7 \\ 
\hline
23 &  -321.7 & -14.2 \\
\hline
24  &  -312.2 &  -14.4 \\
\hline
25 &  -303.6  &  -14.0 \\
\hline
26 &  -294.3 &  -14.3 \\
\end{tabular}
\end{ruledtabular}
\label{tab:gamma_and_lamda}
\end{table}

In the preceding discussion, we have assumed that both the spin-spin coupling and the coupling of the total electronic spin, $\vec{S}$, with the molecular rotation, $\vec{N}$, are negligible.  In this case, the total spin (characterized by $\vec{G}$) is a good quantum number.  However, while these couplings are small, we nevertheless do resolve a splitting of the excited state energy levels by observing three PA resonances as seen in Fig.~\ref{fig:three-features} instead of a single feature for an initial $s$-wave collision.  As we explain later, the ground state is not split in this particular case because $N=0$.

In order to properly label the three PA resonances (associated with ground state $s$-wave collisions) observed for each rovibrational state given spin-spin and spin-rotational coupling,  we redefine $\vec{J}$ to be the total angular momentum \emph{apart from nuclear spin}, $\vec{J}\equiv \vec{N}+\vec{S}$. 
Here, a magnetic coupling between $\vec{S}$ and $\vec{N}$
(involving an interaction term of the form $\hat{H}_{\mathrm{spin-rot}} = \gamma_v \vec{N}\cdot\vec{S}$)
as well as a spin-spin coupling term (of the form $\hat{H}_{\mathrm{spin-spin}} = 2 \lambda_v [\hat{S}_z^2-\hat{S}^2/3]$)
cause a splitting of the rotational levels, previously labeled by $N$, 
according to the $J$ quantum number, given by
$J=(N+S),$ $(N+S-1),$ $(N+S-2)$,$\cdots$, $|N-S|$. Therefore, each level
 with a given $N(\ge S)$ consists of $2S+1$ sub-levels, and the number of sub-levels is equal to the spin multiplicity.
 However, for $N<S$, the number of sub-levels is equal
to $2N+1$ (the rotational multiplicity). Hence, all $N=0$ levels do not split, as mentioned
previously.
For a particular ro-vibrational state, $\left.\left|\nu,N\right.\right>$, with a total spin $S=1$, the rotational energy is  given by~\cite{Herzberg,Schlapp}
%\begin{widetext}
\begin{eqnarray}{\label{SplittingEq1}}
F_{J=N+1}&=&B_vN(N+1)+(2N+3)B_v-\lambda_v\nonumber\\
&-&\sqrt{(2N+3)^2B_v^2+\lambda_v^2-2\lambda_vB_v}+\gamma_v(N+1)\nonumber\\
F_{J=N}~~~&=&B_vN(N+1)\nonumber\\
F_{J=N-1}&=&B_vN(N+1)-(2N-1)B_v-\lambda_v\nonumber\\
&+&\sqrt{(2N-1)^2B_v^2+\lambda_v^2-2\lambda_vB_v}-\gamma_vN,
\end{eqnarray}
%\end{widetext}
where $\lambda_v$ and $\gamma_v$ are constants. Here, $\lambda_v$ is related to the spin-spin interaction and it describes the coupling between the total spin, $\vec{S}$, and the molecular axis;
$\gamma_v$ is related to the spin-rotation interaction and it is a measure of the coupling between $\vec{S}$ and $\vec{N}$.
Under most circumstances, these two constants describe small effects which are not spectroscopically resolvable and are typically ignored in the Dunham expansion. However, at the level of resolution in the current experiment, one needs to take into account these second-order perturbations.
In the case where spin-spin and spin-rotation couplings are small ($B_v \gg |\lambda_v|, \, |\gamma_v|$) we can simplify Eq.~\ref{SplittingEq1} to
\begin{eqnarray}
{\label{SplittingEq2}}
F_{J=N+1}&=&B_vN(N+1)-\frac{2N+2}{2N+3}\lambda_v+\gamma_v(N+1)\nonumber\\
F_{J=N}~~~&=&B_vN(N+1)\nonumber\\
F_{J=N-1}&=&B_vN(N+1)-\frac{2N}{2N-1}\lambda_v-\gamma_vN.
\end{eqnarray}
In addition, when spin-spin coupling is much more important than spin-rotation coupling ($\left|\lambda_v\right|\gg \left|\gamma_v\right|$), the energy ordering results from the $\lambda_v$ terms, and we can label these three peaks in Table \ref{tab:N1-resonances}, energetically from low to high, as $(N'=1, J'=1)$, $(N'=1, J'=2)$, and $(N'=1,  J'=0)$ because $\lambda_v$ is negative.

Using the peak spacings reported in Table \ref{tab:N1-resonances} and Eq.~\ref{SplittingEq2}, we extract the two parameters, $\lambda_v$ and $\gamma_v$.  The determined $\lambda_v$ constants as a function of $v'$ are plotted in Figure~\ref{fig:spinspin}.  The dashed line is provided to show its trend. These results agree well with the previous {\it ab initio} calculation for lithium diatoms~\cite{Minaev2005790}. By using Fig.~3 of Ref.~\cite{Minaev2005790} and averaging $\lambda(R)$ over the internuclear distance $R$ using the wave functions corresponding to the eigenfunctions of the excited state potential curve we refined with our data, we estimate these {\it ab initio} $\lambda_v$ constants for all $v'$ states and plot those also in Fig.~\ref{fig:spinspin}. Note, the uncertainty of the {\it ab initio} results given in Fig. 5 is at least a few tens of MHz.  This results from the estimated error of the original {\it ab initio} calculation (a few percent corresponding to $\approx 10-30$ MHz) and the error ($\approx 10$ MHz) associated with our digitization of the data in Fig.~3 from Ref.~\cite{Minaev2005790}, as well as the fact that the {\it ab initio} calculation was likely done for $^{7,7}$Li$_2$ rather than $^{6,6}$Li$_2$.

This comparison of $\lambda_v$ obtained from experimental data and that obtained from {\it ab initio} calculations clearly demonstrates the validity of the current model to label separate peaks in Table \ref{tab:N1-resonances}.  The values for  $\lambda_v$ and $\gamma_v$ determined from our data are provided in Table \ref{tab:gamma_and_lamda}.  The uncertainty in these parameters is $\pm 400$~kHz and results from the uncertainty in the PA resonance positions.  Using  Eq.~\ref{SplittingEq1}, we verified that the uncertainty in the exact value for $B_v$ does not contribute significantly to the uncertainty in these parameters.  We note that this is the first direct measurement of the spin-spin and spin-rotation coupling constants in a diatomic lithium system.

The interpretation of the PA resonances arising from $p$-wave ground state collisions requires a model to fully account for the hyperfine structure and Zeeman splitting of the ground and excited states at the non-zero magnetic field used. This analysis is the subject for future work.

\section{Refined potential}
\label{sec:thy}

The Morse/Long-Range (MLR) model for potential energy functions was first introduced in 2006 \cite{LeRoy2006,LeRoy2007} and major developments were made in 2009 \citep{LeRoy2009} and 2011 \citep{LeRoy2011}. Over these years, the MLR and related models have been very successful in representing the internuclear potentials for dozens of diatomic and polyatomic molecules \cite{LeRoy2006,LeRoy2007,Salami2007,Shayesteh2007,Li2008,LeRoy2009,Coxon2010,Stein2010,Li2010,Piticco2010,Ivanova2011,LeRoy2011,Dattani2011,Xie2011,Steinke2012}. 

% there's one more paper on a related form, the one where they tried to use p=2

A particular advantage of the MLR model is that it is a single analytic function that very naturally represents the correct shape of the potential function in good agreement with the experimentally observed energies, while also providing an accurate description of the potential's long-range region according to theory. This makes the MLR model especially convenient to use when the known experimental energies are sparse. For example, the most comprehensive study of Li$_2(1^3\Sigma_g^+)$ in existence before this work \cite{Dattani2011}, only had a small amount of data available near the bottom of the potential well, and a small amount at the top (see Figure \ref{fig:globalPotentialAndDataRegions}). In that case, since the MLR model is a single analytic function, it was able to interpolate between those two regions very well (see column 4 of Table \ref{tab:comparisonOfOldAndNewPotential'sPredictions}).

In this section we report a refined version of the Li$_2(1^3\Sigma_g^+)$ MLR potential from \cite{Dattani2011}, that now incorporates the $J=N=1$ energies of the seven new vibrational levels studied in this paper. Various candidate MLR potentials were calculated by Direct-Potential-Fits (DPF), where the parameters of the potential are optimized so that the predicted Schroedinger eigenenergies match the experimental energies as closely as possible. We used the same Hamiltonian form as in \cite{Dattani2011} (see Eqn. 3 of \cite{Dattani2011} and its surroundings), and the least-squares fitting was computed with the publicly available free computer program \texttt{DPotFit} \cite{d:DPotFit}. The candidates for this refined potential were the same type of MLR$_{p,q}^{r_{\rm{ref}}}(N_\beta)$ models for Li$_2(1^{3}\Sigma_{g}^{+})$ as those used in \citep{Dattani2011}.

These models are defined by

\begin{equation}
V_{{\rm MLR}}(r)\equiv\mathfrak{D}_{e}\left(1-\frac{u_{{\rm LR}}(r)}{u_{{\rm LR}}(r_{e})}e^{-\beta(r)\cdot y_{p}^{r_e}(r)}\right)^{2},\label{eq:VMLR}
\end{equation}
where $\mathfrak{D}_{e}$ is the dissociation energy, $r_{e}$ is
the equilibrium internuclear distance, and $\beta(r)$
is

\begin{widetext}
\begin{equation}
\beta(r)\equiv\beta^{r_{\rm{ref}}}_{p,q}(r)\equiv \beta_{\infty}y_{p}^{r_{\rm{ref}}}(r)+\left(1-y_{p}^{r_{\rm{ref}}}(r)\right)\sum_{i=0}^{N_\beta}\beta_{i}\left(y_{q}^{r_{\rm{ref}}}(r)\right)^{i},\label{eq:betaPolynomial}
\end{equation}
\end{widetext}

with 

\begin{equation}
\beta_{\infty}\equiv\lim_{r\rightarrow\infty}\beta(r)=\ln\left(\frac{2\mathfrak{D}_{e}}{u_{{\rm LR}}(r_{e})}\right).
\end{equation}

Eqs. \ref{eq:VMLR} and \ref{eq:betaPolynomial} also depend
on the radial variable

\begin{equation}
y_{n}^{r_{\rm{ref}}}(r)=\frac{r^{n}-r_{{\rm ref}}^{n}}{r^{n}+r_{{\rm ref}}^{n}},
\end{equation}
where the reference distance $r_{{\rm ref}}$ may simply be the equilibrium
distance $r_{e}$ as explicitly written in Eq. \ref{eq:VMLR}, but adjusting it can significantly reduce the required value of $N_\beta$ in Eq. \ref{eq:betaPolynomial} for an accurate fit to the experimental data and more consistent predictions of the physical parameters in the model (see for example Fig. 5 of \cite{Dattani2011}). The \v{S}urkus power $n$ is an integer greater than or equal to 1 which is also adjusted to optimize the fit.

Finally, the definition of $u_{{\rm LR}}(r)$ depends more on the
system being modeled. For large $r$, Eq. \ref{eq:VMLR} usually has
the form $V(r)\simeq\mathfrak{D}_{e}-u_{{\rm LR}}(r)$ \citep{LeRoy2009},
therefore defining $u_{{\rm LR}}(r)$ to be the true theoretical long-range
difference between $\mathfrak{D}_{e}$ and $V(r)$ is often a good
starting choice. As described in \citep{Aubert-Frecon1998}, for very
large internuclear distances, the potential of Li$_{2}$($1^{3}\Sigma_{g}^{+})$
is given by $\mathfrak{D}_{e}-\lambda_{\min}(r)$, so $\lambda_{\min}(r)$
can be a good first choice for $u_{{\rm LR}}(r)$. $\lambda_{\min}(r)$ is defined as the
lowest eigenenergy of the (symmetric) matrix that describes the near-dissociation
interaction of the $1^{3}\Sigma_{g}^{+}$ state with the nearby $1^{1}\Pi_{g}$
and $1^{3}\Pi_{g}$ states. When exchange energy terms and the factors
describing the relativistic retardation effect are neglected, this
matrix is

\begin{widetext}

{\tiny 
\begin{equation}
-\left(\begin{array}{ccc}
D+\frac{1}{3}{\displaystyle \sum_{\substack{m=3,6,8,\\
9,10,11,\ldots
}
}{\frac{C_{m}^{{^{3}\Sigma_{g}^{+}}}+C_{m}^{^{1}\Pi_{g}}+C_{m}^{^{3}\Pi_{g}}}{{r}^{m}}}} & \frac{1}{3\sqrt{2}}{\displaystyle \sum_{\substack{m=3,6,8,\\
9,10,11,\ldots
}
}{\frac{-2C_{m}^{{^{3}\Sigma_{g}^{+}}}+C_{m}^{{^{1}\Pi_{g}}}+C_{m}^{{^{3}\Pi_{g}}}}{{r}^{m}}}} & \frac{1}{\sqrt{6}}{\displaystyle \sum_{\substack{m=3,6,8,\\
9,10,11,\ldots
}
}\frac{-C_{m}^{^{1}\Pi_{g}}+C_{m}^{^{3}\Pi_{g}}}{r^{m}}}\\
\frac{1}{3\sqrt{2}}{\displaystyle \sum_{\substack{m=3,6,8,\\
9,10,11,\ldots
}
}{\frac{-2C_{m}^{{^{3}\Sigma_{g}^{+}}}+C_{m}^{{^{1}\Pi_{g}}}+C_{m}^{{^{3}\Pi_{g}}}}{{r}^{m}}}}\qquad\qquad & D+\Delta E+\frac{1}{6}{\displaystyle \sum_{\substack{m=3,6,8,\\
9,10,11,\ldots
}
}\frac{4C_{m}^{^{3}\Sigma_{g}^{+}}+C_{m}^{^{1}\Pi_{g}}+C_{m}^{^{3}\Pi_{g}}}{r^{m}}} & \frac{1}{2\sqrt{3}}{\displaystyle \sum_{\substack{n=3,6,8,\\
9,10,11,\ldots
}
}\frac{-C_{m}^{^{1}\Pi_{g}}+C_{m}^{^{3}\Pi_{g}}}{{r}^{m}}}\\
\frac{1}{\sqrt{6}}{\displaystyle \sum_{\substack{m=3,6,8,\\
9,10,11,\ldots
}
}\frac{-C_{m}^{^{1}\Pi_{g}}+C_{m}^{^{3}\Pi_{g}}}{r^{m}}} & \frac{1}{2\sqrt{3}}{\displaystyle \sum_{\substack{m=3,6,8,\\
9,10,11,\ldots
}
}\frac{-C_{m}^{^{1}\Pi_{g}}+C_{m}^{^{3}\Pi_{g}}}{{r}^{m}}} & D+\Delta E+\frac{1}{2}{\displaystyle \sum_{\substack{m=3,6,8,\\
9,10,11,\ldots
}
}\frac{C_{m}^{^{1}\Pi_{g}}+C_{m}^{^{3}\Pi_{g}}}{r^{m}}}
\end{array}\right),\label{eq:3x3matrixWithoutExchange}
\end{equation}
}{\tiny \par}

\end{widetext}

\noindent where $\Delta E$ is the (positive) difference between the
$^{2}P_{\nicefrac{1}{2}}$ and $^{2}P_{\nicefrac{3}{2}}$ atomic spin-orbit
levels, $C_3\equiv C_{3}^{^{2S+1}\Lambda_{g}^{(\nicefrac{+}{-})}}$ are the resonance dipole interaction terms, and $C_m\equiv C_{m}^{^{2S+1}\Lambda_{g}^{(\nicefrac{+}{-})}}$ are the dispersion / van der Waals interaction terms.
Note the overall negative sign which reflects that our convention
is the opposite to that in \citep{Aubert-Frecon1998}, where this
matrix was first defined in this form.

Using quantum electrodynamics, it has been shown (see \citep{Meath1968}
and references therein) that the relativistic retardation effect
can be described for large $r$ by making the following modifications
in Eq. \ref{eq:3x3matrixWithoutExchange}:
\begin{widetext}
\begin{align}
C_{3}^{^{3}\Sigma_{g}^{+}} & \rightarrow C_{3}^{^{3}\Sigma_{g}^{+}}\left(\cos\left(\frac{r}{\textrm{\textcrlambda}}\right)+\left(\frac{r}{\textrm{\textcrlambda}}\right)\sin\left(\frac{r}{\textrm{\textcrlambda}}\right)\right)\equiv C_{3,{\rm ret}}^{^{3}\Sigma_{g}^{+}}(r),\\
C_{3}^{^{2S+1}\Pi_{g}} & \rightarrow C_{3}^{^{2S+1}\Pi_{g}}\left(\cos\left(\frac{r}{\textrm{\textcrlambda}}\right)+\left(\frac{r}{\textrm{\textcrlambda}}\right)\sin\left(\frac{r}{\textrm{\textcrlambda}}\right)-\left(\frac{r}{\textrm{\textcrlambda}}\right)^{2}\cos\left(\frac{r}{\textrm{\textcrlambda}}\right)\right)\equiv C_{3,{\rm ret}}^{^{2S+1}\Pi_{g}}(r) ,
\end{align}
\end{widetext}

\noindent where $\textrm{\textcrlambda} = \lambda_{SP} / 2\pi$ is the wavelength of light associated with the atomic $^2S - ^2P$ transition, which for $^6$Li is implicitly given in the caption to Table \ref{tab:parametersForPotential}.

%
%To take into account the weakening of interactions due to the overlap
%of the electronic wavefunctions of the interacting atoms, we also
%replace all $C_{m}$ values in Eq. \ref{eq:3x3matrixWithoutExchange}
%with their damped versions $C_{m}(r)\equiv D_{m}(r)C_{m}$ as described
%in the context of the MLR model in \citep{LeRoy2011}. Following \citep{LeRoy2011},
%we choose for this work the generalized Douketis \emph{et al.} \cite{douketis:3057} damping function:
%
%\begin{equation}
%D_{m}(r)=\left(1-e^{-\left(\frac{b(s)\cdot\rho r}{m}\,+\,\frac{c(s)\cdot(\rho r)^{2}}{m}\right)}\right)^{m+s},
%\end{equation}
%with $s=-1$, where $b(s)$ and $c(s)$ are
%system-independent parameters: $b(-1)=3.30$ and $c(-1)=0.423$.
%Since $^{6,6}$Li$_{2}$ is homonuclear, $\rho=\rho_{{\rm ^{6}Li}}\equiv\left(I_{p}^{{\rm ^{6}Li}}/I_{p}^{{\rm H}}\right)$,
%where $I_{p}^{{\rm A}}$ denotes the ionization potential of an atom
%${\rm A}.$

The only $C_m$ terms used in the analysis of \citep{Dattani2011}
were $C_{3}^{^{3}\Sigma_{g}^{+}},\, C_{3}^{^{1}\Pi_{g}},\, C_{3}^{^{3}\Pi_{g}},\, C_{6}^{^{3}\Sigma_{g}^{+}}$
and $C_{8}^{^{3}\Sigma_{g}^{+}}$ since the $C_{6/8}^{^{2S+1}\Pi_{g}}$
terms were found not to have a significant effect on the $1^{3}\Sigma_{g}^{+}$
potential. Though fairly accurate values for all three $C_{10}$ terms involved were
available (with the infinite mass approximation) as early as 2007
\citep{Zhang2007}, their effects were not considered in \citep{Dattani2011} due to the unavailability
of the values for the three $C_{9}$ terms (which emerge in third-order
perturbation theory) involved. Now that values for the involved $C_{9}$
and $C_{11}$ terms have been reported (with the infinite mass approximation)
in \citep{Tang2011}, we are able to consider all terms appearing
in Eq. \ref{eq:3x3matrixWithoutExchange} up to and including the
$C_{11}$ terms. However, Figure \ref{fig:long-rangePotentialsInLeRoySpace}
shows that none of these $C_9$, $C_{10}$, or $C_{11}$ terms seem to make a noticeable
effect on the potential in our data region, and since including superfluous $C_{m}$ terms might 
lead to unphysical behavior in the short to mid-range regions that
lack data %
\footnote{The parameter $p$ in Eqs. \ref{eq:VMLR} and \ref{eq:betaPolynomial}
must be larger than the difference between $C_{{\rm last}}$ and $C_{{\rm first}}=C_{3}$
(see the supplementary material of  \citep{Dattani2011} for a derivation
of this rule). The larger the value of $p$, the more $\beta_{i}$
parameters that tend to be required in Eq. \ref{eq:betaPolynomial} for
a good fit \citep{LeRoy2007}. When too many $\beta_{i}$ parameters
are used, the potential can become less `rigid' and
may extrapolate poorly where data is not available, which can be problematic in
this study as there are large gaps in the available experimental energies.
},
 we chose not to include these terms, nor the $C_{6/8}^{^{2S+1}\Pi_g}$ terms in our potentials. The $C_m$ values used throughout this paper (including in Figure \ref{fig:long-rangePotentialsInLeRoySpace}) are presented in Table \ref{tab:C_m_constants}, along with details about how each of them was calculated, and their sources. The calculations that were done with the Hylleraas-type basis set are the most accurate; however, no such calculation has been published for the $C_{10}$ values, so for Figure \ref{fig:long-rangePotentialsInLeRoySpace} we used the values from \cite{Zhang2007} which were calculated using the Laguerre-type orbital (LTO) basis. The only values for $C_{9,10,11}$ that have been published were calculated with the approximation that the mass of Li is infinite, while finite mass corrections have been included to calculate $C_{3,6,8}$ for $^6$Li in \cite{Tang2009}. While we use these $^6$Li values for $C_{6,8}$, we use the $^7$Li values for $C_3$ published in  \cite{Tang2010a}, which are likely \cite{Mitroy2012} to be closer to the true $^6$Li values than the $^6$Li values from  \cite{Tang2009} since relativistic corrections have been included in the former. The $C_{6,8,10}$ used for the $a^3\Sigma_u^+$ state were non-relativistic $^6$Li values from \cite{Tang2009} that were calculated with the Hylleraas-type basis set and with finite mass corrections - they are presented in Table \ref{tab:parametersForPotential} and are slightly different from the $^7$Li values used in \cite{Dattani2011}.

The caption to Figure \ref{fig:long-rangePotentialsInLeRoySpace} and the remainder of this paper use the following definitions for notational simplicity:

\begin{equation}
C_m^{\Sigma} \equiv C_m^{^3\Sigma_g^+} 
\end{equation}

\begin{equation}
C_3^\Pi \equiv C_3^{^1\Pi_g} = -C_3^{^3\Pi_g} = \frac{1}{2} C_3^\Sigma
\label{eq:C3PiDefinitionAndSymmetryRelations}.
\end{equation}

\noindent Eq. \eqref{eq:C3PiDefinitionAndSymmetryRelations} also presents useful symmetry relations (see Ref. \cite{Aubert-Frecon1998} and references therein).

\begin{widetext} 
\begin{table}
\caption{$C_m$ coefficients used in our refined potential. Numbers in parenthesis are the uncertainties in the last digits shown.
\label{tab:C_m_constants}}
\begin{tabular}{ c @{~~~} c @{~~~} c @{~~~} c @{~~~} c  @{~~~} c @{~~~} c c}
 & \multirow{2}{*}{Li$_2(1^{3}\Sigma_{g}^{+})$} & \multirow{2}{*}{Li$_2(1^{1}\Pi_{g})$} & \multirow{2}{*}{Li$_2(1^{3}\Pi_{g})$} & \multirow{2}{*}{Basis Set}  & Effective & Relativistic & Source\tabularnewline
& & & & & Mass & Corrections & \tabularnewline
\hline 
\hline 
$C_{3}[\mbox{cm\ensuremath{^{-1}\mbox{\rm{\AA}}^{3}]}}$ & $3.57773 \times 10^5$ & $-1.78887 \times 10^5$ & $1.78887 \times 10^5$ & Hylleraas-type  & $^{7}{\rm Li}$ & \cmark &\cite{Tang2010a}\tabularnewline
$C_{6} [\mbox{cm\ensuremath{^{-1}\mbox{\rm{\AA}}^{6}]}}$ & $1.00059(3) \times 10^7$ & $6.78183(2) \times 10^6$ & $6.78183(2) \times 10^6$ & Hylleraas-type & $^{6}{\rm Li}$ & \xmark  &\cite{Tang2009}\tabularnewline
$C_{8}[\mbox{cm\ensuremath{^{-1}\mbox{\rm{\AA}}^{8}]}}$ & $3.69965(8) \times 10^8$ & $1.39076(1) \times 10^8$ & $6.55441(5) \times 10^8$ & Hylleraas-type & $^{6}{\rm Li}$ &
\xmark & \cite{Tang2009}\tabularnewline
$C_{9}[\mbox{cm\ensuremath{^{-1}\mbox{\rm{\AA}}^{9}]}}$ & $1.6340(1) \times 10^8$ & $3.694(1) \times 10^7$ & $3.694(1) \times 10^7$ & Hylleraas-type &  $^{\infty}{\rm Li}$ & \xmark&\cite{Tang2011}\tabularnewline
$C_{10} [\mbox{cm\ensuremath{^{-1}\mbox{\rm{\AA}}^{10}]}}$ & $1.1374 \times 10^{10}$ & $3.3746 \times 10^9$ & $3.4707 \times 10^8$ & Laguerre-type & $^{\infty}{\rm Li}$ & \xmark & \cite{Zhang2007}\tabularnewline
$C_{11} [\mbox{cm\ensuremath{^{-1}\mbox{\rm{\AA}}^{11}]}}$ & $-1.186 \times 10^{10}$ & $1.985 \times 10^{10}$ & $5.304 \times 10^9$ & Hylleraas-type &  $^{\infty}{\rm Li}$ & \xmark &\cite{Tang2011}\tabularnewline
\end{tabular}
\end{table}
\end{widetext}

Finally, as described in \citep{LeRoy2009,Dattani2011}, the long-range
part of the potential can be modeled more accurately with the inclusion
of another set of adjustments. Following \citep{LeRoy2009,Dattani2011}, we define $C_{6}^{{\rm \Sigma,adj}}\equiv C_{6}^{\Sigma}+\left(C_{3,{\rm ret}}^{\Sigma}\right)^{2}/(4\mathfrak{D}_{e})$
and $C_{9}^{{\rm adj}}\equiv C_{3,{\rm ret}}^{\Sigma}C_{6}^{\Sigma,{\rm adj}}$/(2$\mathfrak{D}_{e})$.  Using these adjustments,
along with the treatment of the retardation effect,
and the removal of the negligible $C_{m}$ terms mentioned in the caption of Figure \ref{fig:long-rangePotentialsInLeRoySpace}, brings us to the definition
of $u_{{\rm LR}}(r)$ we use for this study (which happens to have the same form as that in \cite{Dattani2011}, though the $C_m$ values used are those in Table \ref{tab:C_m_constants} rather than those used in \cite{Dattani2011}):

\begin{equation}
u_{{\rm LR}}(r)=-\lambda_{{\rm min}}^{{\rm adj}}(C_{3,{\rm ret}},C_{6}^{{\rm adj}},C_{8};r)+C_{9}^{{\rm adj}}/r^{9},
\end{equation}
where $\lambda_{{\rm min}}^{{\rm {\rm adj}}}(r)$ is defined as the
lowest eigenenergy of the following matrix $\mathbf{M}_{{\rm LR}}^{{\rm adj}}$
which is a modified version of Eq. \ref{eq:3x3matrixWithoutExchange}:
\begin{widetext}
\begin{equation}
\mathbf{M}_{{\rm LR}}^{{\rm adj}}=\left(\begin{array}{ccc}
-\frac{1}{3}\,\left({\frac{C_{3,{\rm ret}}^{{\Sigma}}}{{r}^{3}}}+{\frac{C_{6}^{{\Sigma,{\rm adj}}}}{{r}^{6}}}+{\frac{C_{8}^{{\Sigma}}}{{r}^{8}}}\right) & \frac{\sqrt{2}}{3}\,\left({\frac{C_{3,{\rm ret}}^{{\Sigma}}}{{r}^{3}}}+{\frac{C_{6}^{{\Sigma,{\rm adj}}}}{{r}^{6}}}+{\frac{C_{8}^{{\Sigma}}}{{r}^{8}}}\right) & \frac{2}{\sqrt{6}}\,\left(\frac{C_{3,{\rm ret}}^{\Pi}}{r^{3}}\right)\\
\frac{\sqrt{2}}{3}\,\left({\frac{C_{3,{\rm ret}}^{{\Sigma}}}{{r}^{3}}}+{\frac{C_{6}^{{\Sigma,{\rm adj}}}}{{r}^{6}}}+{\frac{C_{8}^{{\Sigma}}}{{r}^{8}}}\right) & \Delta E-\frac{2}{3}\,\left(\frac{C_{3,{\rm ret}}^{\Sigma}}{r^{3}}+\frac{C_{6}^{\Sigma,{\rm adj}}}{r^{6}}+\frac{C_{8}^{\Sigma}}{r^{8}}\right) & \frac{1}{\sqrt{3}}\,\left(\frac{C_{3,{\rm ret}}^{\Pi}}{{r}^{3}}\right)\\
\frac{2}{\sqrt{6}}\,\left(\frac{C_{3,{\rm ret}}^{\Pi}}{r^{3}}\right) & \frac{1}{\sqrt{3}}\,\left(\frac{C_{3,{\rm ret}}^{\Pi}}{{r}^{3}}\right) & \Delta E
\end{array}\right).\label{eq:3x3matrixActuallyUsed}
\end{equation}
\end{widetext}

\begin{figure}
\caption{{(color online) Long range potentials in Le$\,$Roy space
demonstrating that all $C_{9},\, C_{10},$ and $C_{11}$
terms appearing in Eq.~\ref{eq:3x3matrixWithoutExchange} do
not contribute significantly to $\lambda_{{\rm min}}$ for the four
vibrational levels between the $m$-dependent Le Roy radius (taken from \cite{Tang2011}) and the
long-range data region, and especially do not contribute much in the
long-range data region. $V_1(r)$ is the experimentally determined potential energy curve from Ref. \cite{Dattani2011}. $V_6(r)$ is the theoretical
non-retarded long-range potential of Aubert-Fr\'{e}con \cite{Aubert-Frecon1998} in which allÊ
$C_m$ coefficients are included from $C_3^{\Sigma,\Pi}$ to $C_{11}^{\Sigma, ^1\Pi_g, ^3\Pi_g}$, at the values given in Table \ref{tab:C_m_constants}.
$V_2(r)$ is the same as $V_6(r)$ but with $C_{6,8,9,10,11}^{^1\Pi_g, ^3\Pi_g}=0$. ÊThis shows thatÊ
the $C_m^{^1\Pi_g, ^3\Pi_g}$ coefficients are unnecessary.
$V_3(r)$ is the same as $V_2(r)$ but also with $C_{11}^{\Sigma}=0$ and shows thatÊ
$C_{11}^{\Sigma}$ is unnecessary.
$V_4(r)$ is the same as $V_3(r)$ but also with $C_{10}^{\Sigma}=0$, which shows that $C_{10}^{\Sigma}$ is unnecessary. Ê$V_5(r)$ is the same as $V_4(r)$Ê
but also with $C_{9}^{\Sigma}=0$, which shows that $C_{9}^{\Sigma}$ is unnecessary.Ê
\label{fig:long-rangePotentialsInLeRoySpace}}}
\includegraphics[width=1\columnwidth]{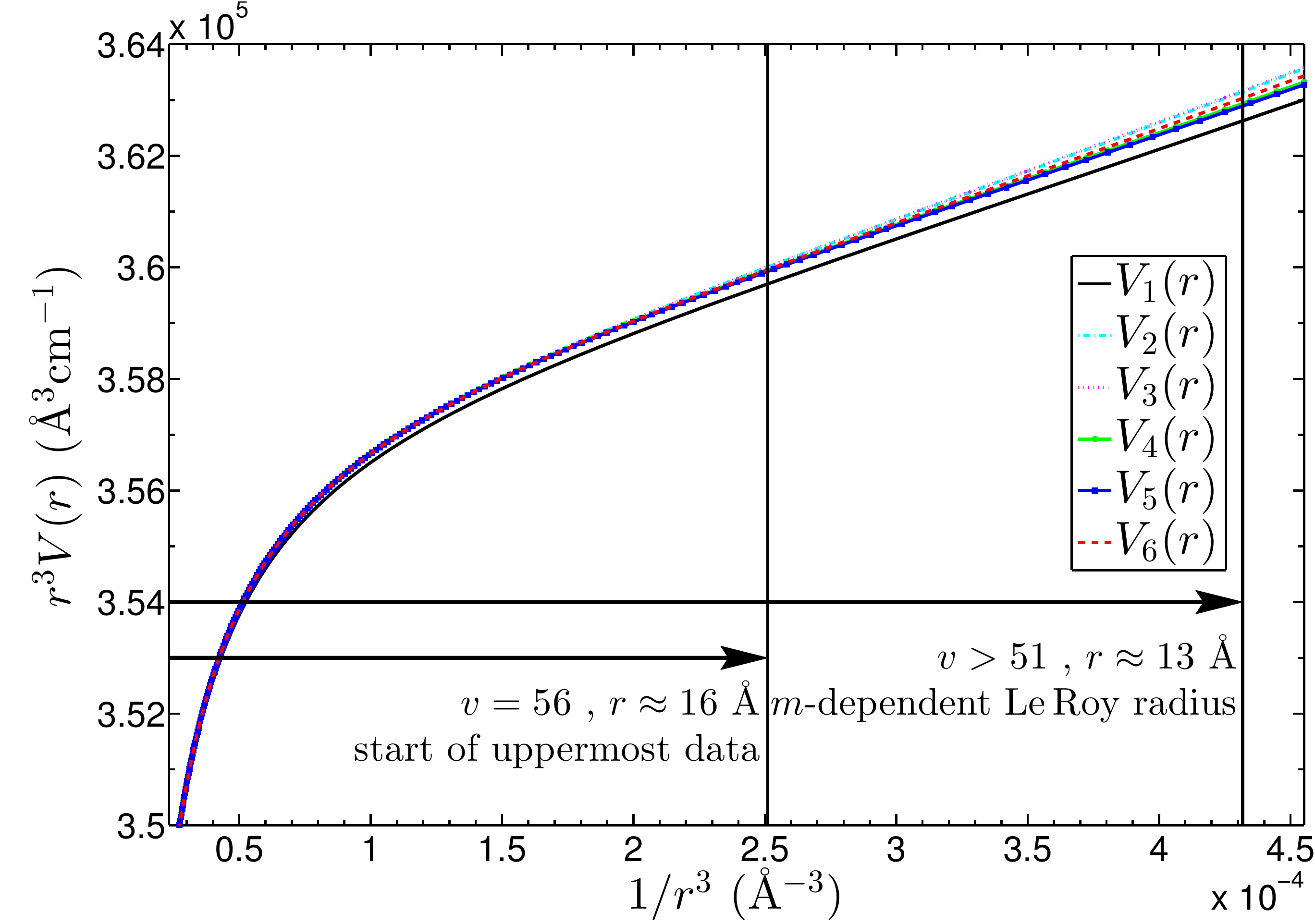}
\end{figure}

\begin{figure}
\caption{{(color online) Long range potentials in Le$\,$Roy space
demonstrating that all $C_{9},\, C_{10},$ and $C_{11}$
terms appearing in Eq.~\ref{eq:3x3matrixWithoutExchange} do
not contribute significantly to $\lambda_{{\rm min}}$ for the four
vibrational levels between the $m$-dependent Le Roy radius (taken from \cite{Tang2011}) and the
long-range data region, and especially do not contribute much in the
long-range data region.
$V_1(r)$ is the experimentally determined potential energy curve from Ref. \cite{Dattani2011}.
$V_2(r)$ is the theoretical long-range potential from Aubert-Frecon in which all 
$C_m$ coefficients are included from $C_3^{\Sigma,\Pi}$ to $C_{11}^{\Sigma, ^1\Pi_g, ^3\Pi_g}$.
$V_3(r)$ is the same as $V_2(r)$ but with $C_{6,8,9,10,11}^{^1\Pi_g, ^3\Pi_g}=0$.  This shows that 
the $C_m^{^1\Pi_g, ^3\Pi_g}$ coefficients are unnecessary.
$V_4(r)$ is the same as $V_3(r)$ but also with $C_{11}^{\Sigma}=0$ and shows that 
$C_{11}^{\Sigma}$ is unnecessary.
$V_5(r)$ is the same as $V_4(r)$ but also with $C_{10}^{\Sigma}=0$, which shows that $C_{10}^{\Sigma}$ is unnecessary.  Finally, $V_6(r)$ is the same as $V_5(r)$ 
but also with $C_{9}^{\Sigma}=0$, which shows that $C_{9}^{\Sigma}$ is unnecessary. 
\label{fig:long-rangePotentialsInLeRoySpace}}}

\includegraphics[width=1\columnwidth]{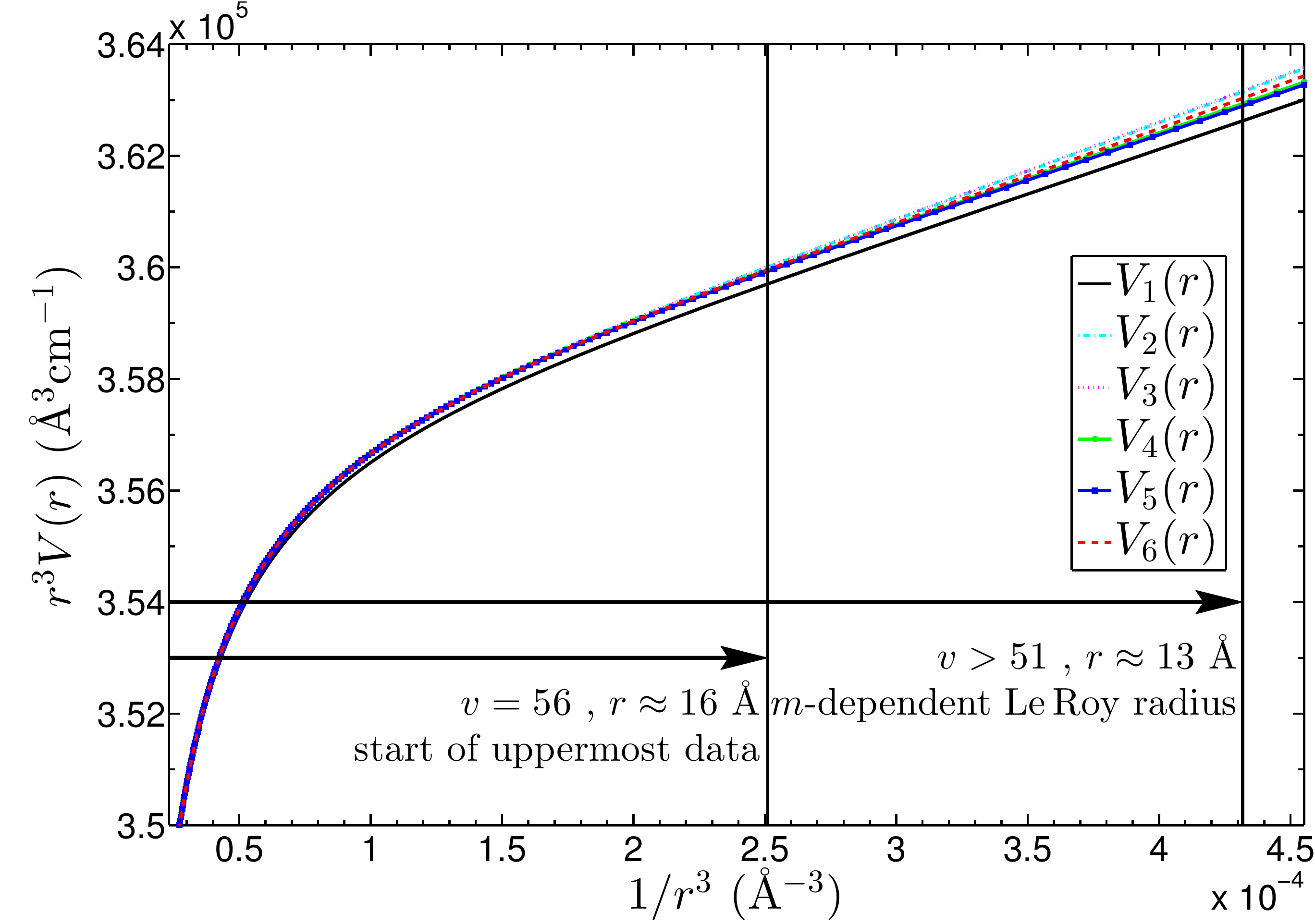}
\end{figure}

To treat more than one isotopologue of Li$_2(1^3\Sigma_g^+)$, we include Born-Oppenheimer Breakdown (BOB) corrections in the same way as described in Section 2.4 and Eqns. 3-7 of \cite{Dattani2011}. Since this study is focused on the $^{6,6}$Li$_2$ isotopologue, we chose to make this the `reference isotopologue' (as defined after Eq. 3 in \cite{Dattani2011}), rather than $^{7,7}$Li$_2$ which was the reference isotopologue in \cite{Dattani2011}.

As in \cite{Dattani2011}, since most of the ro-vibrational observations of the $1^3\Sigma_g^+$ state of Li$_2$ were FTIR emissions from the $1^3\Sigma_g^+$ state into the $a^3\Sigma_u^+$ state \cite{Martin19881369,linton:6036}, our DPF is a multi-state fit to MLR potential models for each of these states. The term values of the relatively small number of observed ro-vibrational energies from PFOODR emissions from the $2^3\Pi_g$ state into the $a^3\Sigma_u^+$ state of $^{7,7}$Li$_2$ \cite{Linton1999} were treated as fitting paramters in the fit, and are provided as supplementary material. 

Since no new data for Li$_2(a^3\Sigma_u^+)$ is being considered, we chose to use the MLR$_{5,3}^{8.0}(3)$-d model for this state, just as in the final potential reported in Table II of \cite{Dattani2011}, except with the parameters re-optimized by the multi-state fit to the current dataset. This is the dataset described in Table I of \cite{Dattani2011}, but expanded to include the $J=N=1$ column (the first column) of Table I of the present paper. The `d' in the name for the MLR model is used to indicate that damping functions are included in the model to take into account the weakening of interactions due to the overlap of the electronic wavefunctions of the interacting atoms as described in \cite{LeRoy2011}. The Li$_2(a^3\Sigma_u^+)$ potentials used for the present work use damping functions that are defined exactly as in Eqns. 12 and 13 of \cite{Dattani2011}; and as in \cite{Dattani2011}, the present paper neglects damping functions for the $1^3\Sigma_g^+$ state for simplicity.

The MLR model chosen in \cite{Dattani2011} for the final Li$_2(1^3\Sigma_g^+)$ potential presented in Table II of that paper was of the form MLR$_{6,3}^{3.8}(9)$. When the parameters of an MLR$_{6,3}^{3.8}(9)$ model are re-optimized by a fit to the current dataset, the resulting potential predicts the ro-vibrational energies very well within their respective experimental uncertainty ranges (on average, over the entire dataset).  However, with the new data (that happens to lie in a region of the potential about which studies previous to this one had no experimental information),  it quickly became apparent that slightly different MLR models could give better fits to the experimental data, and are more robust for predicting the physical parameters $C_3$, $R_e$, and $D_e$. For MLR models different from MLR$_{6,3}^{3.8}(9)$, we obtained starting parameters for the least-squares fitting from the freely available program \texttt{betaFIT} \cite{d:betaFit}. Parameters from \texttt{d:betaFIT} were first re-optimized using \texttt{DPotFit} to the current dataset, but with an upper limit set to $v$. The resulting parameters were then re-optimized with \texttt{DPotFit} to the full current dataset.

\begin{table}
\caption{
Parameters defining our recommended MLR potentials. Parameters
in square brackets were held fixed in the fit, while numbers in parentheses are
95\% confidence limit uncertainties in the last digit(s)
shown. The analysis used the $^{6}$Li $^{2}P_{\nicefrac{1}{2}}\leftarrow ~
^2S_{\nicefrac{1}{2}}$
excitation energy of $\nicefrac{c}{\lambda_{SP}} \equiv$ $D1 = 14903.2967364$
cm$^{-1}$ from \cite{Sansonetti2011}, and the $^{6}$Li
$^{2}P_{\nicefrac{3}{2}}\leftarrow ~ ^2P_{\nicefrac{1}{2}}$
spin-orbit splitting energy of $\Delta E \equiv D2-D1 = 0.3353246$ cm$^{-1}$
with $D2$ from \cite{Brown2013}. Units of length and energy are \AA ~ and
cm$^{-1}$ respectively,
and the polynomial coefficients $\beta_{i}$ are dimensionless.
$\overline{dd}=0.791$.
\label{tab:parametersForPotential}
}
\begin{center}
\[ \begin{array}{ c r@{.}l c r r@{.}l }
\hline\hline
\noalign{\smallskip}
&\multicolumn{2}{c}{a(^3\Sigma_u^+)} && & \multicolumn{2}{c}{c(1\,^3\Sigma^+_g)} \\
\noalign{\smallskip}
\hline
\noalign{\smallskip}
\mathfrak{D}_e  & \multicolumn{2}{c}{\mkern-18mu ~~ 333.7795\,(62)} &~~ & & \multicolumn{2}{c}{\mkern-18mu  ~~~ 7093.4926\,(86)~~~~}  \\                     
r_e                      & 4&170006\,(32)                    & &                      &  3&065436\,(16)  \\
C_6                     &[\,6&7190\times 10^6\,]  & & C_3              &   3&576828\,(44)\times 10^5 \\
C_8                     &[\,1&12635\times 10^8\,] & & C_6^\Sigma & [\,1&00059\times 10^7\,] \\
C_{10}               &[\,2&78694\times 10^9\,] & & C_8^\Sigma & [\,3&69965\times 10^8\,] \\
\rho_{\rm Li}      & \multicolumn{2}{l}{$~[\,0.54\,] $} & & & 
\multicolumn{2}{l}{$~[\,$\infty$\,]$ }  \\
\{p,\,q\}             & \multicolumn{2}{l}{~\{5,\,3\} } 
                        & & & \multicolumn{2}{l}{~\{6,\,2\} }  \\
r_{\rm ref}         & \multicolumn{2}{l}{~[\,8.0\,]} &&& \multicolumn{2}{l}{~[\,4.8\,]} \\
\beta_0               & -0&516129                                             & & &     -1&819413208 \\
\beta_1               & -0&0980                                                & & &      -0&4729259 \\
\beta_2               &  0&1133                                                 & & &      -0&518639 \\
\beta_3               & -0&0251                                                 & & &      -0&16109 \\
\beta_4               & \multicolumn{2}{c}{\mkern-80mu$---$} & & &     -0&8608 \\
\beta_5               & \multicolumn{2}{c}{\mkern-80mu$---$} & & &      3&933 \\
\beta_6               & \multicolumn{2}{c}{\mkern-80mu$---$} & & &      0&965 \\
\beta_7               & \multicolumn{2}{c}{\mkern-80mu$---$} & & &     -2&81 \\
\beta_8               & \multicolumn{2}{c}{\mkern-80mu$---$} & & &     -2&27 \\
\beta_9               & \multicolumn{2}{c}{\mkern-80mu$---$} & & &      1&2 \\
\noalign{\medskip}
\{p_{\rm ad},\,q_{\rm ad}\}  & \multicolumn{2}{l}{~\{6,\,6\}} 
                         & & & \multicolumn{2}{l}{~\{3,\,3\}} \\
u_0              &  0&069\,(12)                              & & &      1&596\,(8) \\
u_1              & \multicolumn{2}{l}{~~~$---$ } & & &  3&1  \,(4) \\
u_2              & \multicolumn{2}{l}{~~~$---$ } & & & -1&5 \,(5)   \\
u_3              & \multicolumn{2}{l}{~~~$---$ } & & & -2&0 \,(5)   \\
u_\infty         & [0&0]                         & & & [1&055740]  \\
\noalign{\medskip}
\hline\hline
\end{array} \]
\end{center}
\end{table}

We focused our studies on MLR$_{p,q}^{r_{\rm{ref}}}(N_\beta)$ models with $p=6$ since $p$ must be larger than 5, and making it 7 or larger may require a much larger $N_\beta$ [56].  Of all the $(p,q)=(6,1-4)$  potentials calculated, the MLR$_{6,2}^{4.8}(9)$ model stood out as having the best balance of reproducing the experimental data closely, maintaining a low value of $N_\beta$, and predicting $C_3$, $R_e$, and $D_e$ values that are consistent with a large number of other MLR models. The MLR$_{6,2}^{r_{\rm{ref}}}(8)$ models were also excellent for $4.5\le r_{\rm{ref}}\le 4.8$, but the predictions of $C_3$, $R_e$, and $D_e$ were not as consistent with respect to $r_{\rm{ref}}$ as they were with higher $N_\beta$ values. While in \cite{Dattani2011} $q<3$ was not considered due to the fact that models with low $q$ values have more of a tendency to result in potentials with inflection points on the inner wall, the added data considered in this study significantly reduced the tendency for inflection points to appear on the inner wall of the $q=3$ potentials, and inflections did not appear here for $q=2$ in the regime surrounding $(N_\beta,r_{\rm{ref}})=(9,4.8)$. For models with $q=1$, it was challenging to find the global minima in the least-squares fitting procedure; and for modest values of $N_\beta$, potentials in our test cases that reproduced the experimental energies more closely with $q=1$ than with $q\in{2,3}$ were not found. Finally, with the $1^3\Sigma_g^+$ state modeled by an MLR$_{6,2}^{4.8}(9)$ function, and the $a^3\Sigma_u^+$ modeled by an MLR$_{5,3}^{8.0}(3)$ function, it was found that the appropriate number of BOB terms did not change from \cite{Dattani2011} (adding two adiabatic or non-adiabitc BOB terms still did not significantly improve the fit to the data for the $1^3\Sigma_g^+$ state, and reducing the number of adiabatic BOB terms by just one for the $1^3\Sigma_g^+$ state had a noteworthy effect on the quality of the fit to the data).

This analysis for choosing the model was done without the sequential rounding and re-fitting (SRR) procedure of \cite{LeRoy1998}.  Once the MLR$_{6,2}^{r_{\rm{ref}}}(9)$ model was chosen, and the appropriate number of BOB terms was chosen, the SRR procedure was applied. First \texttt{DPotFit} was run with all fitting parameters free. Then $C_3$ for the $1^3\Sigma_g^+$ state was manually rounded to the second digit of its 95\% confidence limit uncertainty and \texttt{DPotFit} was re-run with this rounded $C_3$ value fixed and with the rest of the fitting parameters free. Then the $r_e$ values for both states were manually rounded in the same way as $C_3$ was, and \texttt{DPotFit} was re-run with these rounded $r_e$ values fixed and with the remaining fitting parameters free. Finally, $\mathfrak{D}_e$ for both states were rounded manually in the same way as $C_3$ was, and \texttt{DPotFit} was run with these $\mathfrak{D}_e$ values fixed, with all the remaining fitting parameters free, and with \texttt{IROUND}=-1. 

The final potentials for the $a^3\Sigma_u^+$  and $1^3\Sigma_g^+$ states after this SRR procedure are given in Table \ref{tab:parametersForPotential}. The dimensionless root mean square deviation (labeled $\overline{dd}$, and defined by Eq. 2 of \cite{Dattani2011}) of this overall multi-state fit after the SRR procedure was 0.796, which is less than 1\% higher than in \cite{Dattani2011} where $\overline{dd}$ was 0.789, despite the very high accuracy of the new observed levels making the fit much more demanding. The fact that $\overline{dd}<1$ means that the ro-vibrational energies predicted by the fitted potentials match the observed values to well within their experimental uncertainties. In Table \ref{tab:comparisonOfOldAndNewPotential'sPredictions}, the $J=N=1$ energies for the 7 vibrational levels experimentally found in this work are compared to the energies predicted by this new potential, and to those of the potential from Table IIb of the supplementary material of \cite{Dattani2011}, which were calculated without knowledge of these 7 new energies.

Potentials for the $^{7,7}$Li$_2$ isotopologue can be obtained from those for $^{6,6}$Li$_2$ that were presented in Table \ref{tab:parametersForPotential}, by using the BOB corrections as described in Section 2.4 and Eqns. 3-7 of \cite{Dattani2011}; however, since it is simpler to work with potential functions that do not require BOB corrections, a version of Table \ref{tab:parametersForPotential} where the fitting parameters were optimized to the same dataset and with the same MLR models, but with  $^{7,7}$Li$_2$ used as the reference isotopologue, is included as supplementary material for the convenience of those interested primarily in $^{7,7}$Li$_2$. For this, the same SRR procedure was used as for Table \ref{tab:parametersForPotential}. 
%As in \cite{Dattani2011}, the energies predicted by the potentals when $^{7,7}$Li$_2$ is the reference isotopologue are very slightly closer to the experimentally observed energies than when $^{6,6}$Li$_2$ is the reference isotopologue (and this was true even before the SRR procedure) which means that the global minimum for the latter optimization problem is slightly higher than that of the former one, or that the global minimum was not fully reached.

%We used the parameters defined in the $^{6,6}$Li$_{2}$ potential
%published in Table IIb in the supplementary material for \citep{Dattani2011} (except for the changes in the $C_m$ constants described above) as starting parameters for an iterative least-squares fit of all observed
%$^{6,6}$Li$_{2}$ energies published to date (including the seven new
%$J=N=1$ energies reported in Table \ref{tab:N1-resonances} of this paper) to the fitting parameters in Eq. \ref{eq:VMLR}
%and its children.  The resulting parameters for our new recommended potential are given

There is currently a discrepancy between the $1^3\Sigma^+_g$ state $C_3$ value obtained from experiments \cite{LeRoy2009,Dattani2011} and from \textit{ab initio} calculations \cite{Tang2010a} (see also \cite{Tang2011}). The most accurate estimate of the $^{6}$Li $C_3$ value for this state is $3.57773 \times 10^5$cm$^{-1}$\AA$^3$ from \cite{Tang2010a}, which was actually calculated for $^7$Li but is expected to be more accurate than any other currently known estimate for $^6$Li \cite{Mitroy2012} since it was calculated with relativistic effects included. The value of $C_3$ that came from the fit in 2009 \cite{LeRoy2009} of an MLR model potential for the $A(1^1\Sigma_u^+)$ state to experimental data was larger than this \textit{ab initio} value, and the value from the fit in 2011 \cite{Dattani2011} to an MLR potential model for the $c(1^3\Sigma_g^+)$ state was smaller than the \textit{ab initio} value. The new data used in this analysis brought the fitted value of $C_3=(3.57683 \pm 0.00044) \times 10^5$cm$^{-1}$\AA$^3$ from this study closer to the \textit{ab initio} value than that from \cite{Dattani2011}, but it is still significantly smaller than the \textit{ab initio} value, meaning that more data for the $1^3\Sigma_g^+$ state is perhaps required to resolve the current discrepancy between experiment and theory.

\begin{table}
\caption{Comparison of the $J=N=1$ binding energies found experimentally in this work, to the energies predicted by the refined potential in Table \ref{tab:parametersForPotential} and the potential in Table IIb in the supplementary material of \cite{Dattani2011}. All energies are in cm$^{-1}$, and the predicted energies are represented as the predicted (calculated) energy minus the experimental (observed) energy. In our measurements, the initial free atomic state is $2 a_{2s}$ below the hyperfine center of gravity of the $2S_{1/2} + 2S_{1/2}$ threshold (where $a_{2s}$ for $^6$Li is 152.137MHz \cite{Arimondo1977}). Consequently, the binding energy is computed by adding $304.274$~MHz to the $D1$ transition frequency and subtracting our measured frequency for the PA loss feature.
\label{tab:comparisonOfOldAndNewPotential'sPredictions}}
\begin{tabular}{cccc}
\multirow{3}{*}{$v'$} & \multirow{2}{*}{This work} & This work & Ref. \cite{Dattani2011} \\
 &  (experimental) & (predicted) & (predicted) \\
 & &  calc - obs & calc - obs \\
\hline
\hline
$20$ & $2666.12934 \pm 2 \times 10^{-5}$ & $-2.60 \times 10^{-6}$ & $-0.525$ \\
$21$ & $2508.90963 \pm 2 \times 10^{-5}$ & $~ ~ 1.41 \times 10^{-5}$ &  $-0.648$   \\
$22$ & $2357.23922 \pm 2 \times 10^{-5}$ & $-9.80 \times 10^{-7}$ & $-0.781$   \\
$23$ & $2211.13373 \pm 2 \times 10^{-5}$ & $-1.07 \times 10^{-5}$ & $-0.920$  \\ 
$24$ & $2070.60609 \pm 2 \times 10^{-5}$ & $-1.77 \times 10^{-6}$ & $-1.059$ \\
$25$ & $1935.66542 \pm 2 \times 10^{-5}$ & $~ ~1.51 \times 10^{-5}$ & $-1.194$   \\
$26$ & $1806.31575 \pm 2 \times 10^{-5}$ & $-8.91 \times 10^{-6}$ & $-1.319$  \\
\end{tabular}
\end{table}

\section{Conclusions}
\label{sec:conc}

In summary, we have measured the binding energies of seven vibrational levels $v'=20-26$ of the $1^{3}\Sigma_{g}^{+}$ excited state of $^{6}$Li$_2$ molecules with an absolute uncertainty of $\pm 0.00002$~cm$^{-1}$ ($\pm 600$~kHz) by photoassociating a quantum degenerate Fermi gas of lithium atoms.  For each vibrational state, we provide measurements of the three rotational states $N'=0,1,2$.  In addition we observe a splitting of the $N'=1$ excited state due to spin-spin and spin-rotation interactions and we use our data to extract the corresponding interaction constants.  We also use our data to further refine the analytic potential energy function for this state and provide the updated Morse/Long-Range model parameters.  These measurements and refined potential provide a starting point to map the 10 bound levels of the ground triplet state $a ^{3}\Sigma_u^{+}$ of $^{6,6}$Li$_2$ and the 11 bound levels of the $a ^{3}\Sigma_u^{+}$ state of $^{7,7}$Li$_2$ by two-color photoassociation \cite{PhysRevA.68.051403}, and to eventually transfer Feshbach molecules into one of these ground state levels using a two-photon STIRAP (stimulated Raman adiabatic passage) process \cite{K.-K.Ni10102008}.  Molecules in the triplet rovibrational ground state can, in theory, relax to the singlet state $X ^{1} \Sigma_g^{+}$ via inelastic collisions; however, this process, which requires a change of the spin configuration, may be suppressed by weak couplings in the ground state resulting in a spin-blockade metastability of the triplet state.  Such a metastability would be very interesting for the study of ultracold controlled chemistry as ground state triplet molecules possess a magnetic moment and collisions between them should exhibit magnetically tunable scattering resonances \cite{B802322K}.  In addition, work has been done that shows that ground state homo-nuclear molecules with rotational quantum number $N=1$ may be collisionally stable, and these exhibit long-range anisotropic quadrupole-quadrupole interactions \cite{PhysRevA.79.013401}.  Finally, metastable triplet state molecules might be a good candidate for experiments on the alignment and spinning of ultracold molecules  with high-intensity, ultra-short pulsed lasers.

%\begin{acknowledgments}

We gratefully acknowledge Takamasa Momose for the use of the Ti:sapphire laser.  We also thank Robert J Le$\,$Roy for many helpful discussions, and Jim Mitroy for advice on which $C_m$ values to use.  The authors also acknowledge financial support from the Canadian Institute for Advanced Research (CIfAR), the Natural Sciences and Engineering Research Council of Canada (NSERC / CRSNG), and the Canadian Foundation for Innovation (CFI).  N.S.D. also thanks the Clarendon Fund for financial support.

%\begin{thebibliography}{99}
%\bibliographystyle{plain}
%\bibliography{Li_PA}

\end{document}